\documentclass{birkmult}
\usepackage{graphicx}
\def \be{\begin{equation}}
\def \ee{\end{equation}}
\def \bes{\begin{eqnarray}}
\def \ees{\end{eqnarray}}
\begin{document}
%
\title{Quantum Gravity as a quantum field theory of simplicial geometry}
\author{Daniele Oriti}

\address{ Department of Applied Mathematics and Theoretical Physics \\ Centre for Mathematical Sciences, University of Cambridge \\ Wilberforce Road, Cambridge CB3 0WA, England, EU}

\email{d.oriti@damtp.cam.ac.uk}




\begin{abstract}
This is an introduction to the group field theory approach to quantum gravity, with emphasis on motivations and basic formalism, more than on recent results; we elaborate on the various ingredients, both conceptual and formal, of the approach, giving some examples, and we discuss some perspectives of future developments. 
\end{abstract}

\maketitle
\section{Introduction: ingredients and motivations for the group field theory approach}
Our aim in this paper is to give an introduction to the group field theory (GFT) approach to non-perturbative quantum gravity. We want especially to emphasize the motivations for this type of approach, the ideas involved in its construction, and the links with other approaches to quantum gravity, more than reviewing the results that have been obtained up to now in this area. For other introductory papers on group field theory, see \cite{introGFTshort}, but especially \cite{laurentgft}, and for a review of the state of the art see \cite{GFTbook}. No need to say, the perspective on the group field theory approach we provide is a {\it personal} one and we do not pretend it to be shared or fully agreed upon by other researchers in the field, although of course we hope this is the case.
First of all what do we mean by \lq quantizing gravity\rq in the GFT approach? What kind of theory are we after? The GFT approach seeks to construct a theory of quantum gravity that is non-perturbative and background independent. By this we mean that we seek to describe at the quantum level {\it all} the degrees of freedom of the gravitational field and thus obtain a quantum description of the full spacetime geometry; in other words no perturbative expansion around any given gravitational background metric is involved in the definition of the theory, so on the one hand states and observables of the theory will not carry any dependence on such background structure, on the other hand the theory will not include only the gravitational configurations that are obtainable perturbatively starting from a given geometry. Also, let us add a (maybe not necessary) note: we are not after unification of fundamental forces; it cannot be excluded that a group field theory formulation of quantum gravity would be best phrased in terms of unified structures, be it the group manifold used or the field, but it is not a necessary condition of the formalism nor among the initial aims of the approach. So what are group field theories? In a word: group field theories are particular field theories on group manifolds that (aim to) provide a background independent third quantized formalism for simplicial gravity in any dimension and signature, in which both geometry and topology are thus dynamical, and described in purely algebraic and combinatorial terms. The Feynman diagrams of such theories have the interpretation of simplicial spacetimes and the theory provides quantum amplitudes for them, in turn interpreted as discrete, algebraic realisation of a path integral description of gravity. Let us now motivate further the various ingredients entering the formalism (for a similar but a bit more extensive discussion, see \cite{thesis}), and at the same time discuss briefly other related approaches to quantum gravity in which the same ingredients are implemented. 

\subsection{Why path integrals? The continuum sum-over-histories approach}
Why to use a description of quantum gravity on a given manifold in terms of path integrals, or sum-over-histories? The main reason is its generality: the path integral formulation of quantum mechanics, let alone quantum gravity, is more general than the canonical one in terms of states and Hamiltonians, and both problems of interpretation and of recovering of classicality (via decoherence) benefit from such a generalisation \cite{hartle}. Coming to quantum gravity in particular, the main advantages follow from its greater generality: one does not need a canonical formulation or a definition of the space of states of the theory to work with a gravity path integral, the boundary data one fixes in writing it down do not necessarily correspond to canonical states nor have to be of spacelike nature (one is free to consider timelike boundaries), nor the topology of the manifold is fixed to be of direct product type with a {\it space} manifold times a time direction (no global hyperbolicity is required). On top of this, one can maintain manifest diffeomorphism invariance, i.e. general covariance, and does not need any $(n-1) + 1$ splitting, nor the associated enlargement of spacetime diffeomorhism symmetry to the symmetry group of the canonical theory \cite{canodiffeo}. Finally, the most powerful non-perturbative techniques of quantum field theory are based on path integrals and one can hope for an application of some of them to gravity. So how would a path integral for continuum gravity look like? Consider a compact four manifold (spacetime) with trivial topology
$\mathcal{M}$ and all the possible geometries (spacetime metrics up to
diffeomorphisms) that are compatible with it. The partition function of the theory would then be defined \cite{hartle} by an integral over all possible 4-geometries with a
diffeomorphism invariant measure and weighted by a quantum amplitude
given by the exponential of (\lq i \rq times) the
action of the classical theory one wants to quantize, General
Relativity. For computing transition
amplitudes for given boundary configurations of the field, one would
instead consider a manifold $\mathcal{M}$ again, of trivial topology, with two disjoint boundary
components $S$ and $S'$ and given boundary data, i.e. 3-geometries, on
them: $h(S')$ and $h'(S')$, and define the transition amplitude by:
\be
Z_{QG}\left(h(S),h'(S')\right)=\int_{g(\mathcal{M}\mid h(S),h'(S'))}\mathcal{D}g \,e^{i\,S_{GR}(g,\mathcal{M})}
\ee
i.e. by summing over all 4-geometries inducing the given 3-geometries
on the boundary, with the amplitude possibly modified by boundary terms if needed.
The expression above is purely formal: first of all we lack a
rigorous definition of a suitable measure in the space of 4-geometries, second the expectation is that the oscillatory nature of the integrand will make the integral badly divergent. To ameliorate the situation somehow, a \lq Wick rotated\rq of the above expression was advocated with the definition of a \lq\lq Euclidean quantum gravity\rq\rq where the sum would be only over Riemannian metrics with a minus sign in front of the action in the definition of the integral \cite{euclQG}. This however was not enough to make rigorous sense of the theory and most of the related results were obtained in semiclassical approximations \cite{euclQG}. Also, the physical
interpretation of the above quantities presents several challenges,
given that the formalism seems to be bound to a cosmological setting, where our usual interpretations of quantum
mechanics are not applicable. We do not discuss this here, but it is worth keeping this issue in mind, given that a good point about group field theory is that it seems to provide a rigorous version of the above formulas (and much more than that) which is also {\i local} in a sense to be clarified below.

\subsection{Why topology change? Continuum 3rd quantization of gravity}
In spite of the difficulties in making sense of a path integral
quantization of gravity on a fixed spacetime, one can think of doing
even more and treat not only geometry but also topology as a dynamical
variable in the theory. One would therefore try to implement a sort of
\lq\lq sum over topologies\rq\rq alongside a sum over geometries, thus
extending this last sum to run over {\it all} possible spacetime
geometries and not only those that can live on a given topology. Again
therefore the main aim in doing this is to gain in generality: there
is no reason to {\it assume} that the spacetime topology is fixed to
be trivial, so it is good not to assume it. Of course this has
consequences on the type of geometries one can consider, in the
Lorentzian case, given that a non-trivial spacetime topology implies
spatial topology change \cite{fay} and this in turn forces the metric
to allow either for closed timelike loops or for isolated degeneracies
(i.e. the geometry may be degenerate, have zero volume element, at
isolated points). While in a first order or tetrad formulation of
gravity one can thus avoid the first possibility by allowing for the
second, in the second order metric formulation one is bound to include
metrics with causality violations. This argument was made stronger by
Horowitz \cite{horo} to the point of concluding that {\it if}
degenerate metrics are included in the (quantum) theory, then topology
change is not only possible but {\it unavoidable} and non-trivial
topologies therefore {\it must} be included in the quantum
theory. However, apart from greater generality, there are various
results that hint to the {\it need} for topology change in quantum
gravity. Work on topological geons \cite{fayrafael}, topological
configurations with particle-like properties, suggest that spatial
topology change (the equivalent of pair creation for geons) is needed
in order for them to satisfy a generalisation of the spin-statistics
theorem. Work in string theory \cite{greene} indicates that different
spacetime topology can be equivalent with respect to stringy
probes. Wormholes, i.e. spatial topology changing spacetime
configurations, have been advocated as a possible mechanism that turn
off the cosmological constant decreasing its value toward zero
\cite{banks}, and the possibility has been raised that {\it all}
constants of nature can be seen as vacuum {\it parameters}, thus in
principle computed, in a theory in which topology is allowed to
fluctuate \cite{coleman}. This last idea, together with the analogy
with string perturbation theory and the aim to solve some problems of
the canonical formulation of quantum gravity, prompted the proposal of
a \lq\lq third quantization\rq\rq formalism for quantum gravity
\cite{giddingsstrominger, guigan}. The idea is to define
a (scalar) field in superspace $\mathcal{H}$ for a given choice of
basic spatial manifold topology, i.e. in the space of all possible
3-geometries (3-metrics $^3h_{ij}$ up to diffeos) on, say, the 3-sphere, essentially turning the wave function of the canonical
theory into an operator: $\phi(^3h)$, whose dynamics is defined by
an action of the type:
\be
S(\phi)=\int_{\mathcal{H}}\mathcal{D} ^3h\,\phi(^3h)\Delta\phi(^3h) + \lambda \int_{\mathcal{H}}\mathcal{D} ^3h\,\mathcal{V}\left(\phi(^3h)\right)
\ee
with $\Delta$ being the Wheeler-DeWitt operator of canonical
gravity here defining the kinetic term (free propagation) of the
theory, while $\mathcal{V}(\phi)$ is a generic, e.g. cubic,
and generically non-local (in superspace) interaction term for the
field, governing the topology changing processes. Notice that because
of the choice of basic spatial topology needed to define the 3rd
quantized field, the topology changing processes described here are
those turning $X$ copies of the 3-sphere into $Y$ copies of the same. The quantum theory
is defined by the partition function $Z=\int\mathcal{D}\phi 
e^{-S(\phi)}$, that produces the sum over histories outlined above,
including a sum over topologies with definite weights, as a dynamical
process, in its perturbative expansion in Feynman graphs:

\includegraphics[width=7cm]{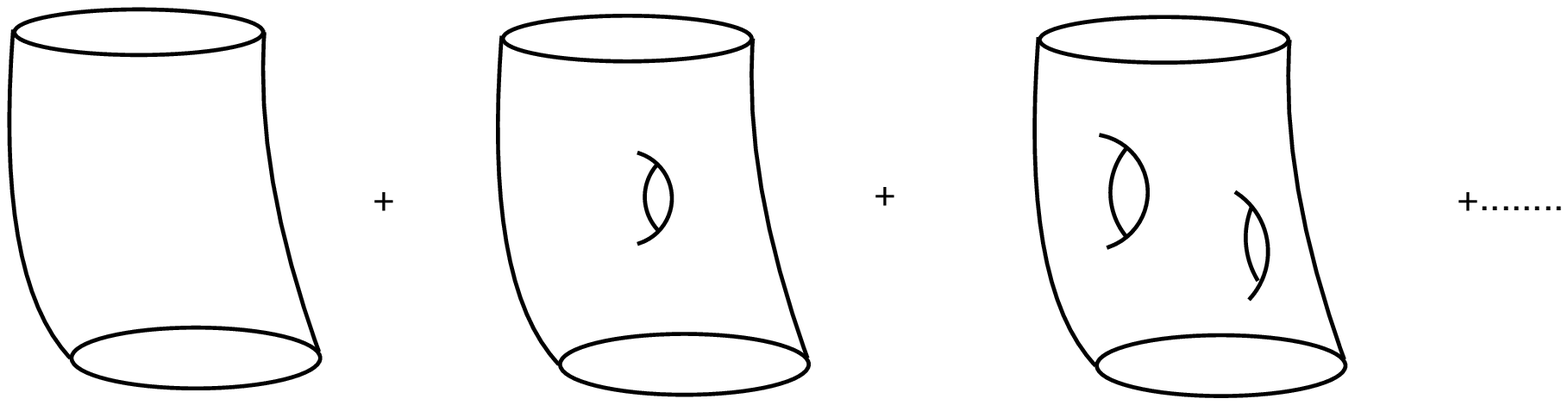}

The
quantum gravity path integral for each topology will represent the Feynman amplitude for each \lq graph\rq, with the one for trivial topology representing a sort
of one particle propagator, thus a Green function for the
Wheeler-DeWitt equation. Some more
features of this (very) formal setting are worth mentioning: 1) the
full classical equations of motions for the scalar field will be a
non-linear extension of the Wheeler-DeWitt equation of canonical
gravity, due to the interaction term in the action, i.e. of the
inclusion of topology change; 2) the perturbative 3rd quantized vacuum
of the theory will be the \lq\lq no spacetime\rq\rq state, and not any
state with a semiclassical geometric interpretation in terms of a
smooth geometry, say a Minkowski state. We will see
shortly how these ideas are implemented in the group field theory
approach.
      
\subsection{Why going discrete? Matrix models and simplicial quantum gravity}
\begin{figure}[t]
\setlength{\unitlength}{1cm}
\begin{minipage}[b]{5.9cm}
\includegraphics[width=5cm, height=2cm]{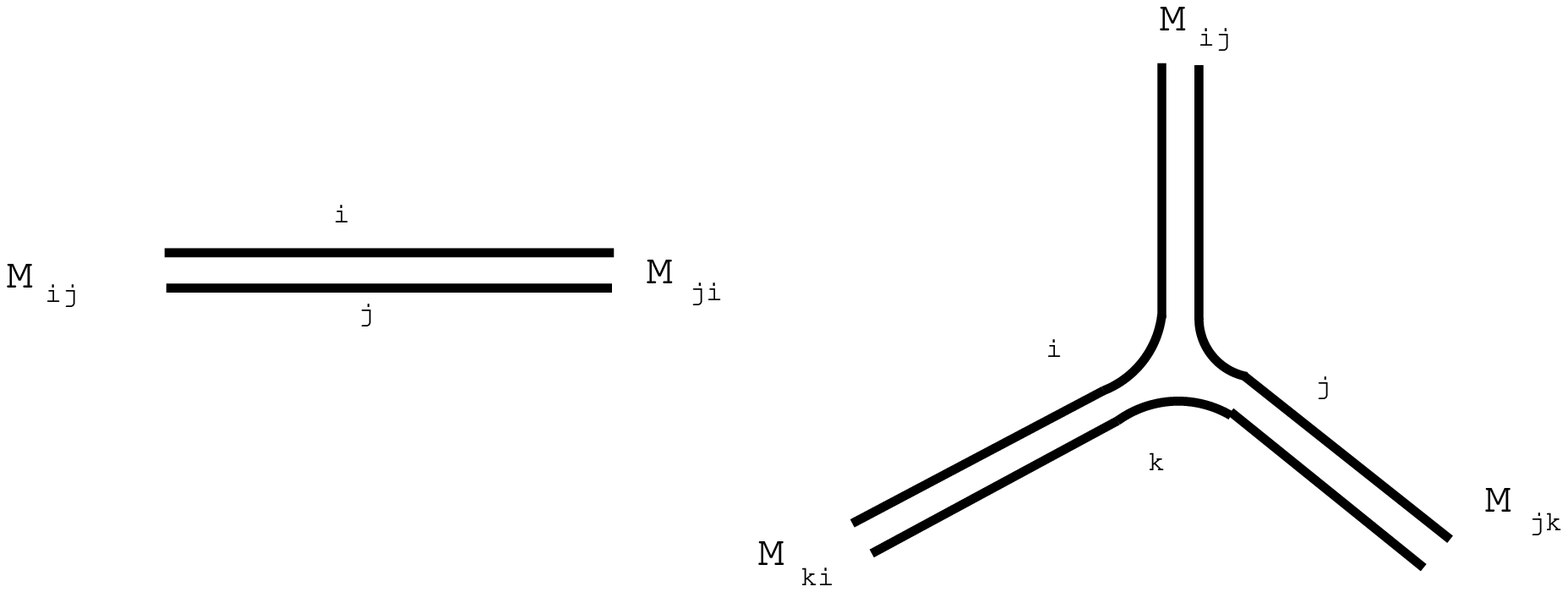}
\caption{\label{fig:matrixpropvertex} Propagator and vertex}
\end{minipage}
\hfill
\begin{minipage}[b]{5.2cm}
\includegraphics[width=5cm, height=2cm]{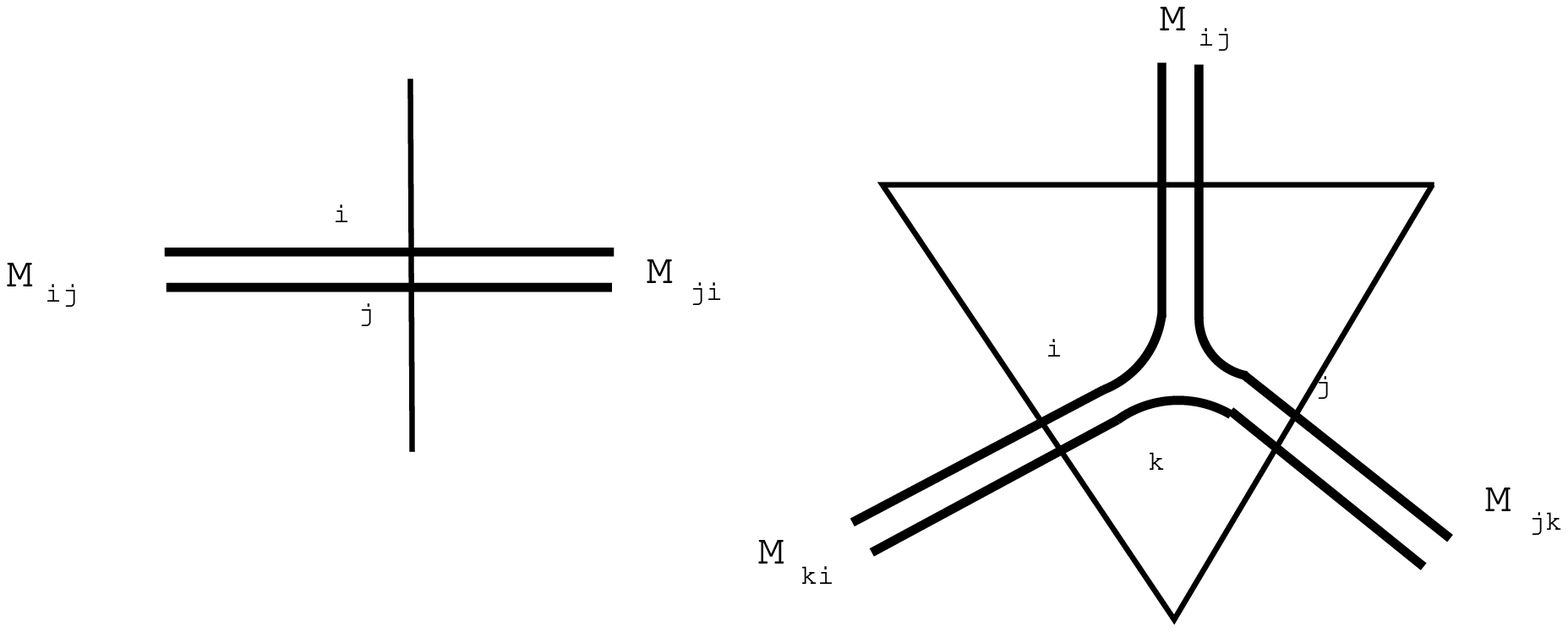}
\caption{\label{fig:matrixdualpropvertex} Dual picture}
\end{minipage}
\end{figure}

However good the idea of a path integral for gravity and its extension
to a third quantized formalism may be, there has been no definite
success in the attempt to realise them rigorously, nor in developing
the formalism to the point of being able to do calculations and then
obtaining solid predictions from the theory. A commonly held opinion
is that the main reason for the difficulties encountered is the use of
a {\it continuum} for describing spacetime, both at the topological
and at the geometrical level. One can indeed advocate the use of {\it
  discrete structures} as a way to regularize and make computable the
above expressions, to provide a more rigorous definition of the
theory, with the continuum expressions and results emerging only in a
continuum {\it limit} of the corresponding discrete quantities.  This
was in fact among the motivations for discrete approaches to quantum
gravity as matrix models, or dynamical triangulations or quantum Regge
calculus. At the same time, various arguments can be and have been put
forward for the point of view that discrete structures instead provide
a more {\it fundamental} description of spacetime. These arguments
come from various quarters. On the one hand there is the possibility,
suggested by various approaches to quantum gravity such as string
theory or loop quantum gravity, that in a more complete description of
space and time there should be a fundamental length scale that sets a
least bound for measurable distances and thus makes the notion of a
continuum loose its physical meaning, at least as a fundamental
entity. Also, one can argue on both philosophical and mathematical
grounds \cite{isham} that the very notion of \lq\lq point\rq\rq can
correspond at most to an idealization of the nature of spacetime due
to its lack of truly operational meaning, i.e. due to the
impossibility of determining with absolute precision the location in
space and time of any event (which, by the way, is implemented
mathematically very precisely in non-commutative models of quantum
gravity, see the contribution by Majid in this volume). Spacetime
points are indeed to be replaced, from this point of view, by small
but finite regions corresponding to our finite abilities in
localising events, and a more fundamental (even if maybe not ultimate
\cite{isham}) model of spacetime should take these local regions as
basic building blocks. Also, the results of black hole thermodynamics
seem to suggest that there should be a discrete number of fundamental
spacetime degrees of freedom associated to any region of spacetime,
the apparent continuum being the result of the microscopic (Planckian)
nature of them. This means that the continuum description of spacetime
will replace a more fundamental discrete one as an {\it approximation}
only, as the result of a {\it coarse graining} procedure. In other
words, a finitary topological space \cite{sorkin} would constitute a
better model of spacetime than a smooth manifold. All these arguments
against the continuum and in favor of a finitary substitute of it can
be naturally seen as arguments in favor of a simplicial description of
spacetime, with the simplices playing indeed the role of a finitary
substitute of the concept of a point or fundamental event, or of a
minimal spacetime region approximating it.   
Simplicial approaches to quantum gravity are matrix models, dynamical
triangulations and quantum Regge calculus. The last one \cite{ruth} is
the straightforward translation of the path integral idea in a
simplicial context. One starts from the definition of a discrete
version of the Einstein-Hilbert action for General Relativity on a
simplicial complex $\Delta$, given by the Regge action $S_R$ in which
the basic geometric variables are the lengths of the edges of
$\Delta$, and then defines the quantum theory usually via Euclidean
path integral methods, i.e. by:
\bes
Z(\Delta)\,=\,\int \mathcal{D}l\;e^{-\,S_R(l)}.
\ees 
The main issue is the definition of the integration measure for the edge lengths, since it has to satisfy the 
discrete analogue of the diffeomorphism invariance of the continuum theory (the most used choices are the $l dl$ and the
$dl/l$ measures) and then the proof that the theory admits a good continuum limit in which continuum general relativity is recovered, indeed the task that has proven to be the most difficult.
Matrix models \cite{matrix} can instead be seen as a surprisingly
powerful implementation of the third quantization idea in a simplicial
context, but in an admittedly simplified framework: 2d Rieammian
quantum gravity. Indeed group field theories are a generalisation of
matrix models to higher dimension and to Lorentzian signature.
Consider the action
\be
S(M)=\frac{1}{2}tr{M^2}\,-\,\frac{\lambda}{3!\sqrt{N}}\,tr{M^3}
\ee
for an $N\times N$ hermitian matrix $M_{ij}$, and the associated
partition function $Z=\int dM e^{-S(M)}$. This in turn is expanded in
perturbative expansion in Feynman diagrams; propagators and vertices
of the theory can be expressed diagrammatically
\ref{fig:matrixpropvertex}, and the corresponding Feynman diagrams,
obtained as usual by gluing vertices with propagators, are given by
{\it fat graphs} of {\bf all} topologies. Moreover, propagators can be
understood as topologically dual to edges and vertices to triangles
\ref{fig:matrixdualpropvertex} of a 2-dimensional simplicial complex
that is dual to the whole fat graph in which they are combined; this
means that one can define a model for quantum gravity in 2d, via the
perturbative expansion for the matrix model above, as sum over {\bf
  all 2d triangulations} $T$ of {\bf all topologies}. 
Indeed the
amplitude of each Feynman diagram for the above theory is related to
the Regge action for classical simplicial gravity in 2dm for fixed
edge lengths equal to $N$ and positive cosmological constant, and more
specifically, the partition function is:
\be
Z=\int dM e^{-S(M)}=\sum_{T} \frac{1}{sym(T)}\lambda^{n_2(T)}N^{\chi(T)}
\ee   
where $sym(T)$ is the order of symmetries of the triangulation $T$, $n_2$ is the number of triangles in it, and $\chi$ is the Euler characteristic of the same triangulation.
Many results have been obtained over the years for this class of
models, for which we refer to the literature \cite{matrix}. Closely related to matrix models is the dynamical triangulations approach \cite{DT}, that extends the idea and results of defining a path integral for gravity as a sum over equilateral triangulations of a given topology to higher dimensions, weighted by the (exponential of the) Regge action for
gravity:
\be
Z(G,\lambda,a)=\sum_{T}\frac{1}{sym(T)}e^{iS_R(T,G,\Lambda,a))}
\ee
where $G$ is the gravitational constant and $\Lambda$ is a
cosmological constant. In the Lorentzian case one also distinguishes
between spacelike edges (length square $a^2$) and timelike ones
(length square $-a^2$), and imposes some additional restrictions on
the topology considered and on the way the triangulations are
constructed via the gluing of d-simplices. In particular, one may then
look for a continuum limit of the theory, corresponding to the limit
$a\rightarrow 0$ accompanied by a suitable renormalisation of the
constants of the theory $\Lambda$ and $G$, and check whether in this
limit the structures expected from a continuum quantum gravity theory
are indeed recovered, i.e. the presence of a smooth phase with the
correct macroscopic dimensionality of spacetime.
And indeed, the exciting recent results obtained in this approach seem
to indicate that, in the Lorentzian context and for trivial topology,
a smooth phase with the correct dimensionality is obtained even in 4
dimensions, which makes the confidence in the correctness of the strategy
adopted to define the theory grow stronger.

\subsection{Why groups and representations? Loop quantum gravity and spin foams}
We will see many of the previous ideas at work in the group field
theory context. There, however, a crucial role is played by the
Lorentz group and its representations, as it is in terms of them that
geometry is described. Another way to see group field theories in fact
is as a re-phrasing (in addition to a generalisation) of the matrix
model and simplicial quantum gravity formalism in an algebraic
language. Why would one want to do this? One reason is the physical
meaning and central role that the Lorentz group plays in gravity and
in our description of spacetime; another is that by doing this, one
can bring in close contact with the others yet another approach to
quantum gravity: loop quantum gravity, through spin foam models. But
let us discuss one thing at the time. The Lorentz group enters
immediately into play and immediately in a crucial role as soon as one
passes from a description of gravity in terms of a metric field to a
first order description in terms of tetrads and connections. Gravity
becomes not too dissimilar from a gauge theory, and as such its basic
observables (intended as correlations of partial observables
\cite{carlobook}) are given by parallel transports of the connection
itself along closed paths, i.e. holonomies, contracted in such a way
as to be gauge invariant. Indeed these have a clear operational
meaning \cite{carlobook}. The connection field is a $so(3,1)$ valued
1-form (in 4d) and therefore its parallel transports define elements
of the Lorentz group, so that the above observables (in turn
determining the data necessary to specify the states of a canonical
formulation of a theory based on this variables) are basically given
by collections of group elements associated to possible paths in
spacetime organized in the form of networks. They are {\it classical
  spin networks}. In a simplicial spacetime, the valence of these
networks will be constrained but they will remain the basic
observables of the theory. A straightforward quantization of them
would be obtained by the choice of a representation of the Lorentz
group for each of the links of the network to which group elements are
associated. Indeed, the resulting quantum structures are {\it spin
  networks}, graphs labeled by representations of the Lorentz group
associated to their links, of the type characterizing states and
observables of loop quantum gravity \cite{carlobook}, the canonical
quantization of gravity based on a connection formulation. A covariant
path integral quantization of a theory based on spin networks will
have as histories (playing the role of a 4-dimensional spacetime
geometries) a higher-dimensional analogue of them: a spin foam
\cite{review,alex,thesis}, i.e. a 2-complex (collection of faces
bounded by links joining at vertices) with representations of the
Lorentz group attached to its faces, in such a way that any slice or
any boundary of it, corresponding to a spatial hypersurface, will be
indeed given by a spin network. Spin foam models \cite{review, alex,
  thesis} are intended to give a path integral quantization of gravity based on
these purely algebraic and combinatorial structures.  

In most of the current models the combinatorial structure of the spin
foam is restricted to be topologically dual to a simplicial complex of
appropriate dimension, so that to each spin foam 2-complex it
corresponds a simplicial spacetime, with the representations attached
to the 2-complex providing geometric information to the simplicial
complex; in fact they are interpreted as volumes of the
(n-2)-simplices topologically dual to the faces of the 2-complex. The
models are then defined
by an assignment of a quantum probability amplitude (here factorised
in terms of face, edge, and vertex contributions)to each spin foam $\sigma$
summed over, depending on the representations $\rho$ labeling it, also
being summed over,
i.e. by the transition amplitudes for given boundary spin networks
$\Psi,\Psi'$ (which may include the empty spin network as well):
$$
Z=\sum_{\sigma\mid\Psi,\Psi'}w(\sigma)\sum_{\{\rho\}}\prod_{f}A_{f}(\rho_f)\prod_{e}A_{e}(\rho_{f\mid
  e})\prod_{v}A_{v}(\rho_{f\mid v}); $$
one can either restrict the sum over spin foams to those
corresponding to a given fixed topology or try to implement a sum over
topologies as well; the crucial point is in any case to come up with a
well-motivated choice of quantum amplitudes, either coming from some
sort of discretization of a classical action for gravity or from some
other route. Whatever the starting point, one would then have an implementation of a sum-over-histories for gravity in a
combinatorial-algebraic context, and the key issue would then be to
prove that one can both analyse fully the quantum domain, including
the coupling of matter fields, and at the same recover classical and
semi-classical results in some appropriate limit. 
\begin{figure}[t] 
\setlength{\unitlength}{1cm}
\begin{minipage}[b]{5.357cm}
\includegraphics[width=4cm, height=4cm]{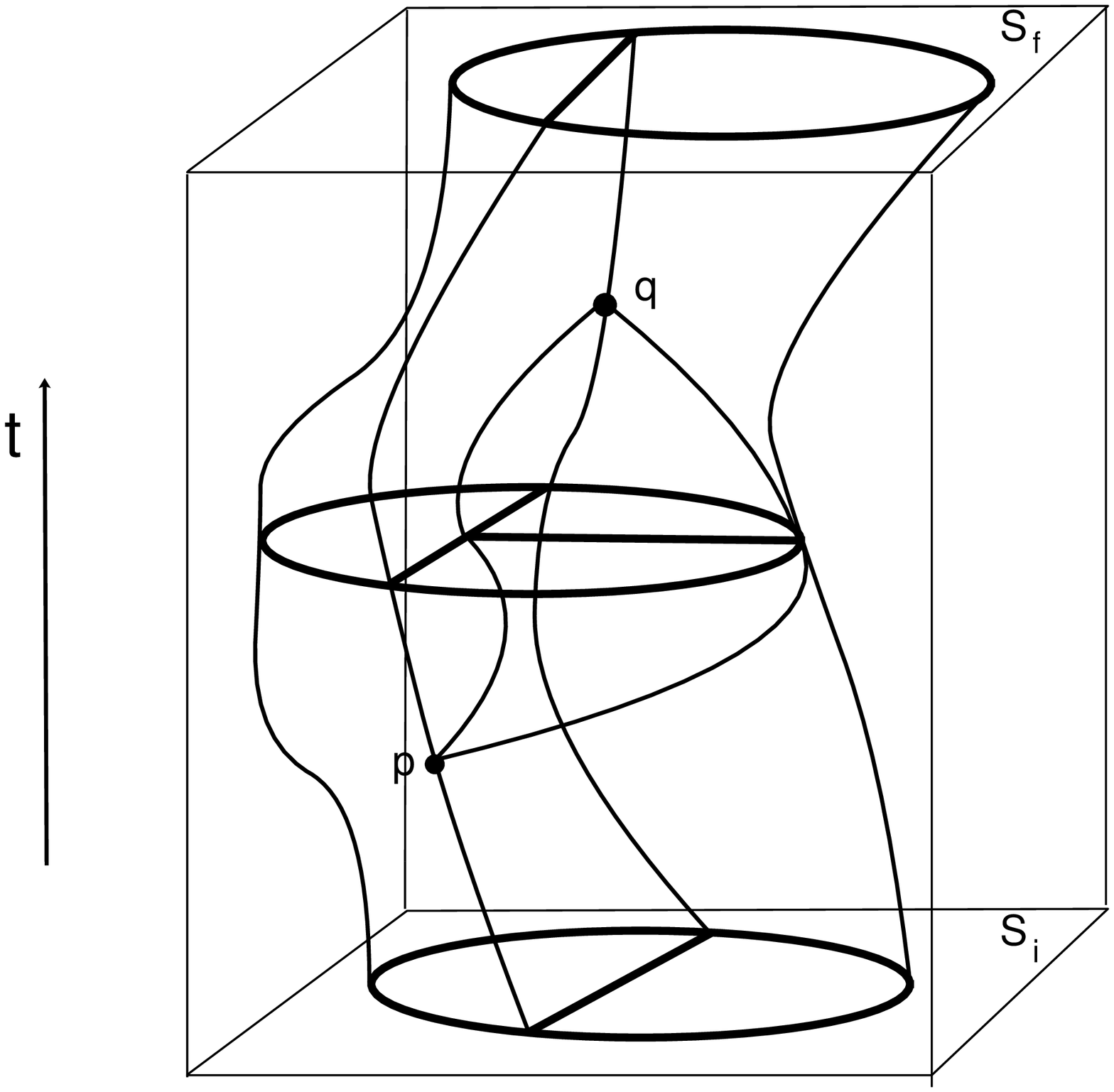}
\caption{\label{fig:spinfoam} A spin foam}
\end{minipage}
\hfill
\begin{minipage}[b]{5.278cm}
\includegraphics[width=3.5cm, height=2.5cm]{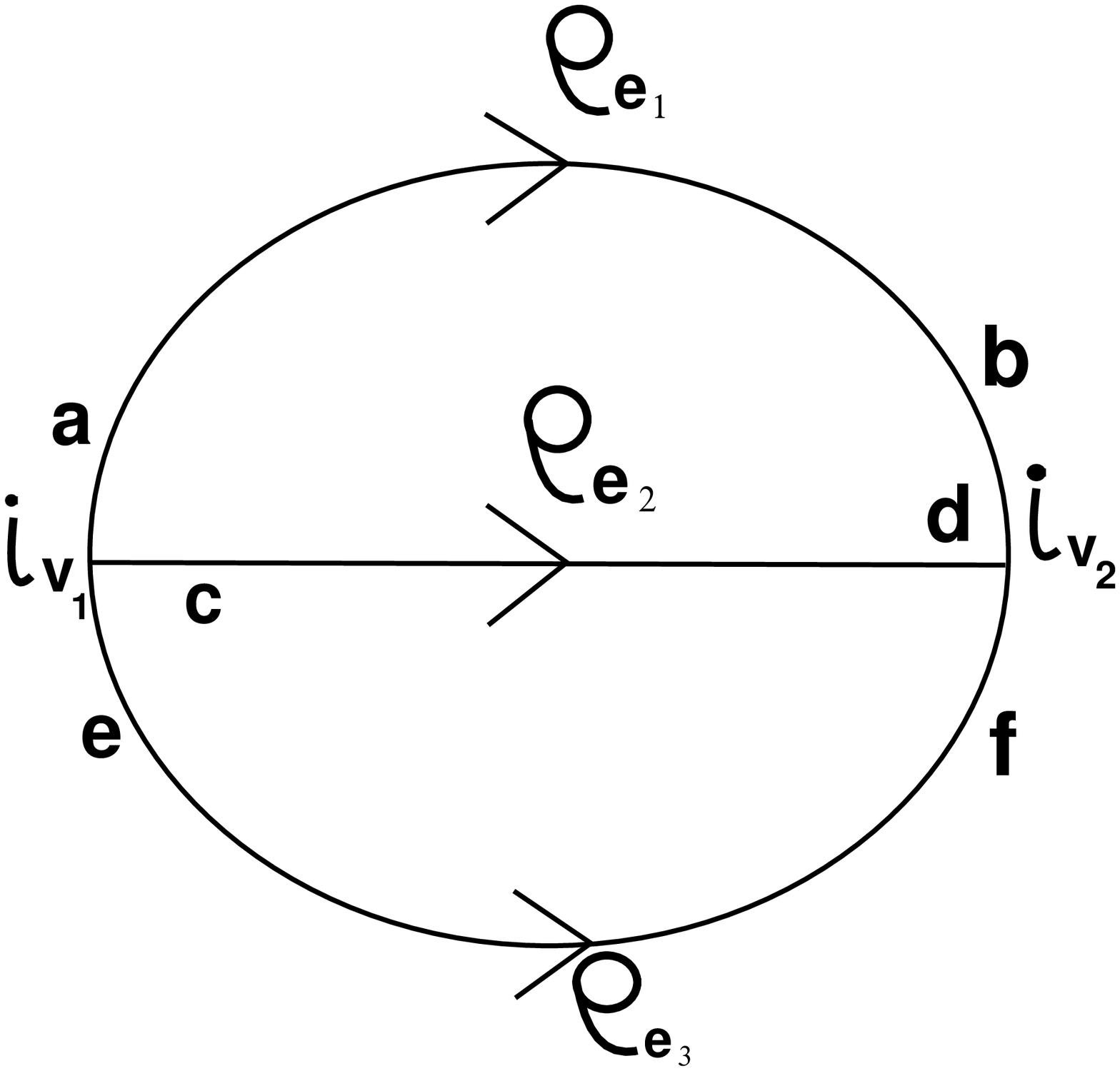}
\caption{A spin network}
\end{minipage}
\end{figure}
A multitude of results have been already
obtained in the spin foam approach, for which we refer to \cite{review,alex,thesis}.
We will see shortly that this version of the path integral idea is the one coming out naturally from group field theories.

\section{Group field theory: what is it? The basic GFT formalism}
Group field theories, as anticipated, are a new realization of the third quantization idea that we have outlined above, in a simplicial setting, and in which the geometry of spacetime as well as superspace itself are described in an algebraic language. 
As such, they bring together most of the ingredients entering the other approaches we have briefly discussed, thus providing hopefully a general encompassing framework for developing them, as we will try to clarify in the following. We describe the basic framework of group field theories and the rationale for its construction first, and then we will give an explicit (and classic) example of it so to clarify the details of the general picture.
 
\subsection{A discrete superspace}
The first ingredient in the construction of a third quantization
theory of gravity in $n$ dimensions is a definition of superspace,
i.e. the space of (n-1)-geometries. In a simplicial setting, spacetime
is discretized to a simplicial complex and thus it is built out of
fundamental blocks represented by $n$-simplices; in the same way, an
(n-1)-space, i.e. an hypersurface (not necessarily spacelike) embedded
in it, is obtained gluing together along shared (n-2)-simplices a
number of (n-1)-simplices in such a way as to reproduce through their
mutual relations the topology of the hypersurface. In other words, a
(n-1)-space is given by a (n-1)-dimensional triangulation and its
geometry is given not by a metric field (thanks to which one can
compute volumes, areas, lengths and so on) but by the geometric data
assigned to the various elements of the complex: volumes, areas,
lengths etc. There is some freedom in the choice of variables to use
as basic ones for describing geometry and from which to compute the
various geometric quantitities. In Regge calculus, as we have seen,
the basic variables are chosen to be the edge lengths of the complex;
in group field theories \cite{laurentgft,carlomike}, as currently
formulated, the starting assumption is that one can use as basic
variables the volumes of (n-2)-simplices (edge lengths in 3d, areas of
triangles in 4d, etc). The consequences and possible problems
following from this assumption have not been fully investigated
yet. These (n-2)-volumes are determined by unitary irreducible
representations $\rho$ of the Lorentz group, one for each $(n-2)$ face
of the simplicial complex. Equivalently, one can take as basic
variables appropriate Lorentz group elements $g$ corresponding to the
parallel transports of a Lorentz connection along dual paths (paths
along the cell complex dual to the triangulation), one for each
(n-2)-face of the complex. The equivalence between these two sets of
variables is given by harmonic analysis on the group, i.e. by a
Fourier-type relation between the representations $\rho$ and the group
elements $g$, so that they are interpreted as conjugate variables, as
momenta and position of a particle in quantum mechanics
\cite{carlomike}. Therefore, if we are given a collection of
(n-1)-simplices together with their geometry in terms of associated
representations $\rho$ or group elements $g$, we have the full set of
data we need to characterize our superspace. Now one more assumption
enters the group field theory approach: that one can exploit the
discreteness of this superspace in one additional way, i.e. by
adopting a {\bf local} point of view and considering as the
fundamental superspace a single (n-1)-simplex; this means that one
considers each (n-1)-simplex as a \lq\lq one-particle state\rq\rq, and
the whole (n-1)-d space as a \lq\lq multiparticle\rq\rq state, but
with the peculiarity that these many \lq\lq particles\rq\rq (many
(n-1)-simplices) can be glued together to form a collective extended
structure, i.e. the whole of space. The truly fundamental superspace
structure will then be given by a single (n-1)-simplex geometry,
characterized by $n$ Lorentz group elements or $n$ representations of
the Lorentz group, all the rest being reconstructed from it, either by
composition of the fundamental superspace building blocks (extended
space configurations) or by interactions of them as a dynamical
process (spacetime configurations), as we will see. In the generalised
group field theory formalism of \cite{generalised}, one uses an
extended or parametrised formalism in which additional variables
characterize the geometry of the fundamental (n-1)-simplices, so that
the details of the geometric description are different, but the
overall picture is similar, in particular the local nature of the
description of superspace is preserved.
             
\subsection{The field and its symmetries}
Accordingly to the above description of superspace, the fundamental
field of GFTs, as in the continuum a scalar field living on it,
corresponds to the 2nd quantization of a (n-1)-simplex. The 1st
quantization of a 3-simplex in 4d was studied in detail in \cite{bb}
in terms of the algebraic set of variables motivated above, and the
idea is that the field of the GFT is obtained promoting
to an operator the wave function arising from the 1st quantization of
the fundamental superspace building block.  
We consider then a complex scalar field over the tensor product of $n$
copies of the Lorentz group in $n$ dimensions and either Riemannian or
Lorentzian signature, $$\phi(g_1,g_2,...,g_n): G^{\otimes n}
\rightarrow \mathbb{C}.$$ The order of the arguments in the field,
each labeling one of its $n$ boundary faces ((n-2)-simplices),
corresponds to a choice of orientation for the geometric (n-1)-simplex
it represents; therefore it is natural to impose the field to be
invariant under even permutations of its arguments (that do not change
the orientation) and to turn into its own complex conjugate under odd
permutations. This ensures \cite{DP-P} that the Feynman graphs of the
resulting field theory are given by orientable 2-complexes, while the
use of a real field, with invariance under any permutation of its
arguments, has as a result Feynman graphs including non-orientable
2-complexes as well. If the field has to correspond to an
(n-1)-simplex, with its $n$ arguments corresponding to an
(n-2)-simplex each, one extra condition is necessary: a global gauge
invariance condition under Lorentz transformations \cite{bb}. We thus
require the field to be invariant under the global action of the
Lorentz group, i.e. under the simultaneous shift of each of its $n$
arguments by an element of the Lorentz group, and we impose this
invariance through a projector operator:
$P_g\phi(g_1;
g_2;...;g_n)=\int_G
dg\,\phi(g_1g;g_2g;...;g_ng)$\footnote{The Lorentzian case, with the use of the non-compact Lorentz group as symmetry group, will clearly involve, in the defintion of the symmetries of the field as well as in the definition of the action and of the Feynman amplitudes, integrals over a non-compact domain; this produces trivial divergences in the resulting expressions and care has to be taken in making them well-defined. However, this can be done quite easily in most cases with appropriate gauge fixing. We do not discuss issues of convergence here in order to simplify the presentation.}. Geometrically, this imposes that the
$n$ (n-2)-simplices on the boundary of the (n-1)-simplex indeed close
to form it \cite{bb}; algebraically, this causes the field to be
expanded in modes into a linear combination of Lorentz group invariant
tensors (intertwiners). The mode expansion of the field takes in fact
the form:
$$
\phi^\alpha(g_i,s_i)=\sum_{J_i,\Lambda,k_i}\phi^{J_i\Lambda}_{k_i}\prod_i
D^{J_i}_{k_il_i}(g_i)C^{J_1..J_4\Lambda}_{l_1..l_4}, $$ with the $J$'s
being the representations of the Lorentz group, the $k$'s vector
indices in the representation spaces, and the $C$'s are intertwiners
labeled by an extra representation index $\Lambda$. 
In the generalised formalism of \cite{generalised}, the Lorentz group is extended to $(G\times \mathbb{R})^n$ with consequent extension of the gauge invariance one imposes and modification of the mode expansion.
Note also that the timelike or spacelike nature of the (n-2)-simplices
corresponding to the arguments of the field depends on the group
elements or equivalently to the representations associated to them,
and nothing in the formalism prevents us to consider timelike
(n-1)-simplices thus a superspace given by a timelike (n-1)-geometry. 

\subsection{The space of states or a third quantized simplicial space}
The space of states resulting from this algebraic third quantization
is to have a structure of a Fock space, with N-particle states created
out of a Fock vacuum, corresponding as in the continuum to the \lq\lq
no-spacetime\rq\rq state, the absolute vacuum, not possessing any
spacetime structure at all. Each field being an invariant tensor under
the Lorentz group (in momentum space), labeled by $n$ representations
of the Lorentz group, it can be described by a $n$-valent spin network
vertex with $n$ links incident to it labeled by the
representations. One would like to distinguish a \lq creation \rq and
an \lq annihilation\rq part in the mode expansion of the field, as
$\phi^{J_i\Lambda}_{k_i} = \varphi^{J_i\Lambda}_{k_i} +
\left(\varphi^{J_i\Lambda}_{k_i}\right)^\dagger$, and then one would
write something like:
$\varphi^{J_i\Lambda}_{k_i} \mid 0\rangle$ for a one particle state, $\varphi^{J_i\Lambda}_{k_i} \varphi^{\tilde{J}_i\tilde{\Lambda}}_{\tilde{k}_i} \mid 0\rangle$ for a {\it disjoint} 2-particle state (two disjoint (n-1)-simplices), or 
$\varphi^{J_1 J_2..J_n\Lambda}_{k_1 k_2...k_n} \varphi^{\tilde{J}_1
  J_2...\tilde{J}_n\tilde{\Lambda}}_{\tilde{k}_1 k_2...\tilde{k}_n}
\mid 0\rangle$ for a {\it composite} 2-particle state, made out of two
(n-1)-simplices glued along one of their boundary (n-2)-simplices (the
one labeled by $J_2$), and so on. Clearly the composite states will
have the structure of a spin network of the Lorentz group. This way
one would have a Fock space structure for a third quantized simplicial
space of the same type as that of usual field theories, albeit with
the additional possibility of creating or destroying at once composite
structures made with more than one fundamental \lq quanta\rq of
space. At present this has been only formally realised \cite{mikovic}
and a more complete and rigorous description of such a third quantized
simplicial space is needed.

\subsection{Quantum histories or a third quantized simplicial spacetime}
In agreement with the above picture of (possibly composite) quanta of
a simplicial space being created or annihilated, group field theories
describe the evolution of these states in perturbation theory as a
scattering process in which an initial quantum state (that can be
either a collection of disjoint (n-1)-simplices, or spin network
vertices, or a composite structure formed by the contraction of
several such vertices, i.e. an extended (n-1)-dimensional
triangulation) is transformed into another one through a process
involving the creation or annihilation of a number of quanta. Being
these quanta (n-1)-simplices, their interaction and evolution is
described in terms of $n$-simplices, as fundamental interaction
processes, in which $D$ (n-1)-simplices are turned into $n+1-D$ ones
(in each n-simplex there are n+1 (n-1)-simplices). Each of these fundamental interaction processes corresponds to a possible n-dimensional Pachner move, a sequence of which is known to allow the transformation of any given (n-1)-dimensional triangulation into any other. A
generic scattering process involves however an arbitrary number of
these fundamental interactions, with given boundary data, and each of
these represents a possible quantum history of simplicial geometry, so
our theory will appropriately sum over all these histories with
certain amplitudes. The states being collections of suitably
contracted spin network vertices, thus spin networks themselves
labeled with representations of the Lorentz group (or equivalently by
Lorentz group elements), dual to triangulations of a (n-1)-dimensional
space, their evolution history will be given by 2-complexes labeled
again by representations of the Lorentz group, dual to $n$-dimensional
simplicial complexes. Spacetime is thus purely virtual in this
context, as in the continuum third quantized formalism and as it
should be in a sum over histories formulation of quantum gravity, here
realised as a sum over labeled simplicial complexes or equivalently
their dual labeled complexes, i.e. spin foams. We see immediately
that we have here a formalism with the ingredients of the other
discrete and algebraic approaches to quantum gravity we have outlined
above.   

\subsection{The third quantized simplicial gravity action}
The action of group field theories \cite{laurentgft,introGFTshort,review, alex,generalised} is defined so to implement the above ideas, and it is given by:
 \begin{eqnarray*} S_n(\phi, \lambda)= \frac{1}{2}\prod_{i=1,..,n}\int
  dg_id\tilde{g}_i
  \phi(g_i)\mathcal{K}(g_i\tilde{g}_i^{-1})\phi(\tilde{g}_i) +
  \frac{\lambda}{n+1}\prod_{i\neq j =1}^{n+1}\int dg_{ij}
  \phi(g_{1j})...\phi(g_{n+1 j})\,\mathcal{V}(g_{ij}g_{ji}^{-1}),
\end{eqnarray*}
where of course the exact choice of the kinetic and interaction
operators is what defines the model. We see that indeed the
interaction term in the action has the symmetries and the
combinatorial structure of a $n$-simplex made out of $n+1$
(n-1)-simplices glued pairwise along common (n-2)-simplices,
represented by their arguments, while the kinetic term represent the
gluing of two $n$-simplices along a common (n-1)-simplex, i.e. the
free propagation of the (n-1)-simplex between two
interactions. $\lambda$ is a coupling constant governing the strength
of the interactions, and the kinetic and vertex operators satisfy the
invariance property $\mathcal{K}(g_i\tilde{g}_i^{-1})=
\mathcal{K}(gg_i\tilde{g}_i^{-1}g')$ and
$\mathcal{V}(g_{ij}g_{ji}^{-1})=\mathcal{V}(g_ig_{ij}g_{ji}^{-1}g_j^{-1})$
as a consequence of the gauge invariance of the field itself. A
complete analysis of the symmetries of the various group field theory
actions has not been carried out yet, and in 3d for example it is
known that there exist symmetries of the Feynman amplitudes (i.e. of
the histories) of the theory that are not yet identified at the level
of the GFT action. In the generalised models \cite{generalised}, the
structure of the action is exactly the same, with the group extended
to $G\times\mathbb{R}$. The simplest choice of action is given by
$\mathcal{K}=\int dg\prod_{i=1}^n \delta(g_ig \tilde{g}_i^{-1})$ and
$\mathcal{V}=\prod_{i=1}^{n+1}\int dg_{i} \prod_{i < j}
\delta(g_{ij}g_ig_j^{-1}g_{ji}^{-1})$, that corresponds to a GFT
formulation of topological BF theories in $n$ dimensions, that gives
gravity in 1st order formalism in 3d, as we will see shortly, while in
dimension n=2 gives a sum over matrix models of increasing matrix
dimension if one choses $SU(2)$ as group manifold
\cite{laurentgft}. Less trivial actions can be constructed
\cite{generalised}, while a simple modification of the BF action gives
much studied models of 4d quantum gravity \cite{review,
  alex}. Unfortunately, we do not understand much at present of the
classical theory described by these actions, and paradoxically we
understand better the (perturbative) quantum theory, thanks to the
work done in the context of spin foam models \cite{review, alex}, and
we turn now to this.

\subsection{The partition function and its perturbative expansion}
The partition function of the theory is then given by an integral over
the field of the exponential of (minus) the GFT action. Our current
understanding of the non-perturbative properties of the partition
function, i.e. of the quantum theory, is quite poor, even if some work is currently
in progress on instantonic calculations in GFTs \cite{instantons}. More is known abour perturbative dynamics in terms of Feynman graphs, thanks to work on spin foam models. The
perturbation expansion of the partition function is as usual given by
the Schwinger-Dyson expansion in Feynman graphs:
$$ Z\,=\,\int
\mathcal{D}\phi\,e^{-S[\phi]}\,=\,\sum_{\Gamma}\,\frac{\lambda^N}{sym[\Gamma]}\,Z(\Gamma),
$$
where $N$ is the number of interaction vertices in the Feynman graph
$\Gamma$, $sym[\Gamma]$ is a symmetry factor for the graph and
$Z(\Gamma)$ the corresponding Feynman amplitude.
The Feynman amplitudes can be constructed easily after identification
of the propagator, given by the inverse of the kinetic term in the
action, and the vertex amplitude of the theory; each edge of the
Feynman graph is made of $n$ strands running parallel to each other,
one for each argument of the field, and each is then re-routed at the
interaction vertex, with the combinatorial structure of an
$n$-simplex. Diagrammatically:  
\begin{figure}[here]
\setlength{\unitlength}{1cm}
\begin{minipage}[b]{5.9cm}
\includegraphics[width=5cm, height=2cm]{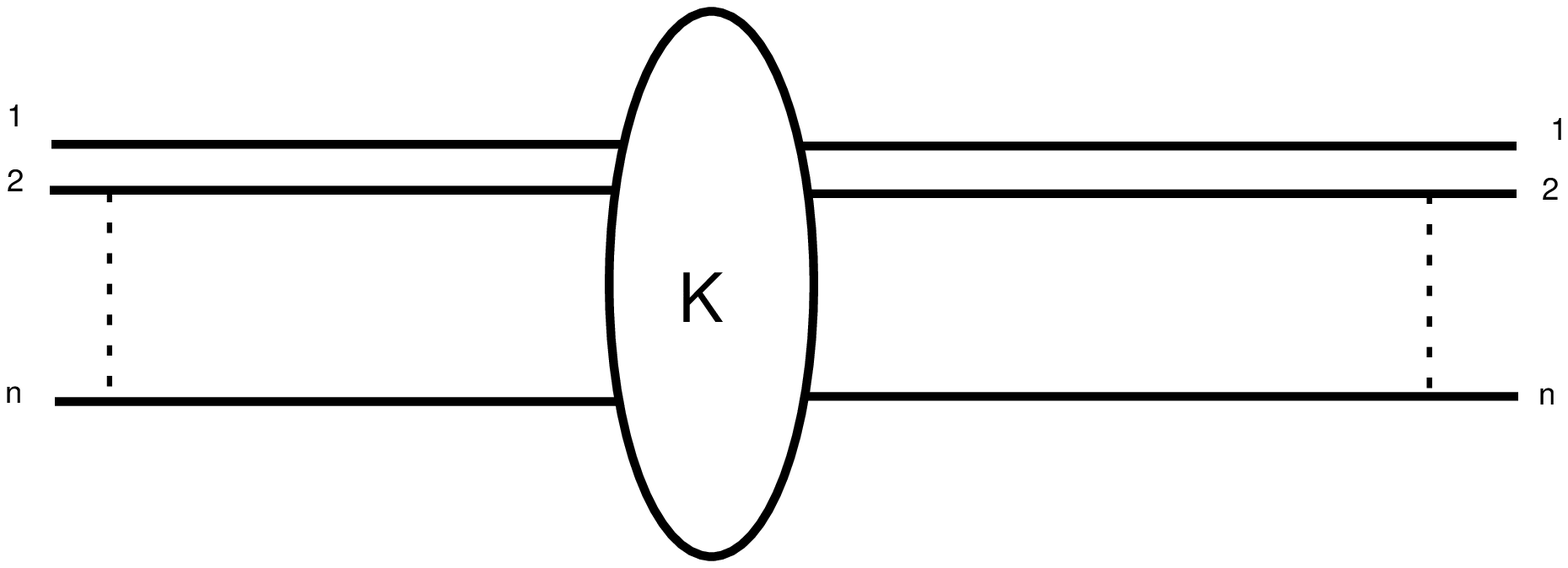}
\end{minipage}
\hfill
\begin{minipage}[b]{5.2cm}
\includegraphics[width=5cm, height=3cm]{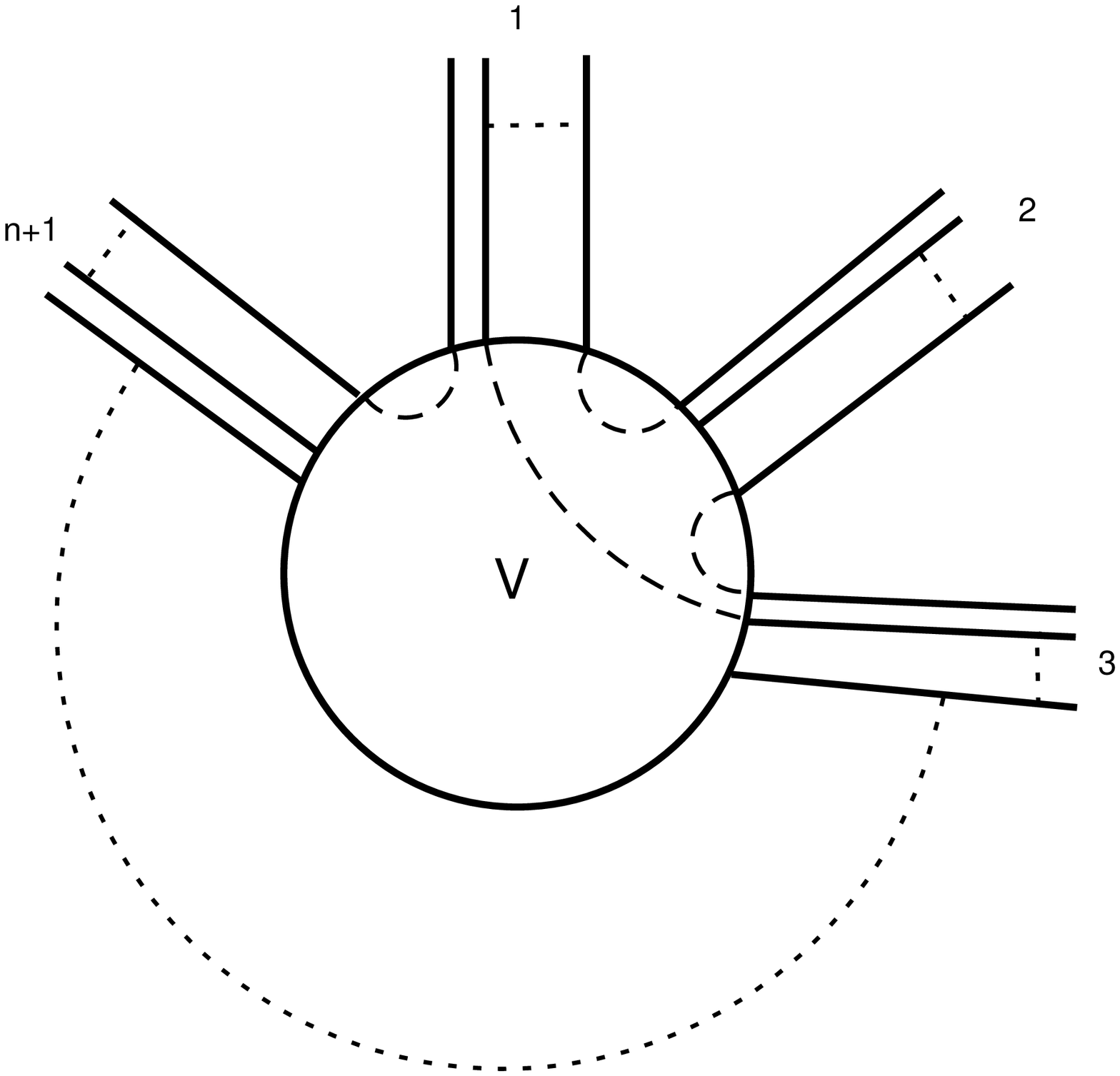}
\end{minipage}
\end{figure}

Each strand in an edge of the Feynman graph goes through several
vertices and then comes back where it started, for closed Feynman
graphs, and therefore identifies a 2-cell. The collection of 2-cells
(faces), edges and vertices of the Feynman graph then characterizes a
2-complex that, because of the chosen combinatorics for the arguments
of the field in the action, is dual to a $n$-dimensional simplicial
complex. Each strand carries a field variable, i.e. a group element in
configuration space or a representation label in momentum
space. Therefore in momentum space each Feynman graph is given by a
spin foam, and each Feynman amplitude (a complex function of the
representation labels) of the GFT by a spin foam model. Indeed, one
can show that the inverse is also true: any local spin foam model can
be obtained from a GFT perturbative expansion
\cite{carlomike,laurentgft}. The sum over Feynman graphs for the
partition function gives then a sum over spin foams (histories of the
spin networks on the boundary in any scattering process considered),
and equivalently a sum over triangulations, augmented by a sum over
algebraic data (group elements or representations) with a geometric
interpretation. This is true of course also for the generalised GFT
models of \cite{generalised}. This perturbative expansion of the
partition function also allows (in principle) the explicit evaluation
of expectation values of GFT observables; these are given
\cite{laurentgft} by gauge invariant combinations of the field
operators that can be constructed using spin networks. In particular,
the transition amplitude (probability amplitude for a certain
scattering process) between certain boundary data represented by two
spin networks can be expressed as the expectation value of field
operators contracted as to reflect the combinatorics of the two spin
networks \cite{laurentgft}.  

\subsection{GFT definition of the canonical inner product}
Even though, as mentioned in the beginning, this is not meant to be a review of the results obtained so far in the GFT approach, for which we refer to \cite{GFTbook}, we would like to mention just one important recent result, because it shows clearly how useful this new formalism can be in addressing long-standing open problems of quantum gravity research. Also, it proves that the overall picture of GFTs that we have outline above (and will summarize in the next subsection) is consistent and fertile. In ordinary QFT, the {\it classical} equations of motion for the (free) 2nd quantized field are also the {\it quantum} equation of motion for the 1st quantized relativistic particle wave function; in other words, the full dynamical content of the 1st quantized (non-interacting) theory is contained in the classical level of the 2nd quantized theory. If we buy the picture of GFTs as representing a 3rd quantization formalism of (simplicial) gravity, we expect a similar situation in which the classical GFT level encodes all the dynamical information about canonical quantum gravity (with fixed topology). We will mention in the final part of this paper some caveats to this perspective and some open issues. Given the limits in our understanding of the classical structure of GFT and of the non-perturbative level of the theory, at present we may hope to see these ideas realised explicitely only in the perturbative expansion of the partition function. Indeed, the hopes are fullfilled. It has been shown \cite{laurentgft} that the restriction of the sum over Feynman graphs outlined above to {\it tree level}, thus neglecting all quantum correction and encoding the classical information only, gives a definition of the \lq 2-point function\rq, for fixed boundary spin networks, that is positive semidefinite and finite, including a sum over all triangulations and thus fully triangulation independent; the triangulations involved are of fixed trivial topology, as appropriate for dealing with canonical quantum gravity, and the sum over geometric data (representations) is well-defined. This prompts the interpretation of the resulting quantity as a well-posed and computable definition of the canonical inner product between quantum gravity states, represented by spin networks, the object that in canonical loop quantum gravity encodes the whole dynamical content of the theory (the action of the Hamiltonian constraint operator on spin network states). Even if more work is certainly needed to build up on this result, we see that the use of GFT techniques and ideas has an immediate and important usefulness even from the point of view of canonical quantum gravity: the GFT definition of the physical inner product for canonical spin network states provides finally a {\it solution} to the long standing issues of canonical quantum gravity on the definition of the Hamiltonian constraint operator, its action on kinematical states, the definition of the physical inner product trough some kind of projection operator formalism, the computation of physical observables, etc. Now it is time to put this concrete proposal to test. Finally, let us stress that this results also highlights the richness of the GFT formalism, and suggests that we have barely started to scratch the surface, with much more lying underneath. If the classical level of GFTs encodes already the full content of canonical quantum gravity, it is clear that a complete analysis of the quantum level of the GFTs will lead us even much further, in understanding a quantum spacetime, than we have hoped to do by studying canonical quantum gravity. this includes the physics of topology change, of course, and the use of quantum gravity sum over histories for other puroses than defining the canonical inner product, but probably much more than that. 

\subsection{Summary: GFT as a general framework for quantum gravity}
Let us briefly summarise the nature of GFTs, before giving an explicit
example of it, so to clarify the details of the formalism. We have a
field theory over a group manifold that makes no reference to a
physical spacetime (except implicitly in the combinatorial structure
that one chooses for the GFT action), in which the field (thus the
fundamental \lq\lq particle\rq\rq it describes) has a geometric
interpretation of a quantized (n-1)-simplex. The states of the theory
(given in momentum space by spin networks) are correspondingly
interpreted as triangulations of (n-1)-dimensional (pseudo)manifolds
\cite{DP-P}. The theory can be dealt with perturbatively through an
expansion in Feynman graphs that describe the possible interactions of
the field quanta, that can be created and annihilated as well as
change intrinsic geometry (configuration variables and associated momenta), as a scattering
process. Geometrically each possible interaction process for given
boundary states is a possible simplicial spacetime with assigned
geometry, and it is given by a spin foam, to which the theory assigns
a precise quantum amplitude. The simplicial spacetimes summed over
have arbitrary topology, as are constructed by all possible gluings of
fundamental interaction vertices each corresponding to an
$n$-simplex. GFTs are therefore a {\it third quantized formulation of
  simplicial geometry}. 
Interestingly, group field
theories also have all the ingredients that enter other approaches to
quantum gravity: boundary states given by spin networks, as in loop
quantum gravity, a simplicial description of spacetime and a sum over
geometric data, as in Quantum Regge calculus, a sum over
triangulations dual to 2-complexes, as in dynamical triangulations, a
sum over topologies like in matrix models, of which GFTs are indeed
higher-dimensional analogues, as we said, an ordering of fundamental events
(vertices of Feynman diagrams), given by the orientation of the
2-complex, which has similarities to that defining causal
sets \cite{CS}, and quantum amplitudes for histories that are given by spin foam models. Therefore one can envisage GFTs as a {\it general
framework for non-perturbative quantum gravity}, that encompasses most
of the current approaches. This is at present only a vague and quite
optimistic point of view, not yet established nor strongly supported
by rigorous results, but we feel that it can be a fruitful point of
view both for the development of GFTs themselves and of these other
approaches as well, by offering a new perspective on them and possibly
new techniques that can be used to address the various open issues
they still face.

\section{An example: 3d Riemannian Quantum Gravity}
For simplicity we consider explicitly in more detail only the 3d Riemannian quantum
gravity case, whose group field theory formulation was first given by
Boulatov \cite{boulatov}.
Consider the real field: $\phi(g_1,g_2,g_3): (SU(2))^{3} \rightarrow
\mathbb{R}$, with the symmetry: $\phi(g_1 g, g_2 g, g_3 g) =
\phi(g_1,g_2,g_3)$, imposed through the projector: $\phi(g_1,g_2,g_3)
= P_g \phi(g_1,g_2,g_3) = \int dg \,\phi(g_1 g,g_2 g,g_3 g)$ and the
symmetry: $\phi(g_1,g_2,g_3) =
\phi(g_{\pi(1)},g_{\pi(2)},g_{\pi(3)})$, with $\pi$ an arbitrary
permutation of its arguments, that one can realise through an explicit
sum over permutations: $\phi(g_1,g_2,g_3) =
\sum_{\pi}\phi(g_{\pi(1)},g_{\pi(2)},g_{\pi(3)})$. In this specific
case, the interpretation is that of a 2nd quantized triangle with its
3 edges corresponding to the 3 arguments of the field; the irreps of
$SU(2)$ labeling these edges in the mode expansion of the field have
the interpretations of edge lengths. The classical
theory is defined by the action:
\begin{eqnarray*} S[\phi] &=& \frac{1}{2}\int
dg_1..dg_3 [P_g\phi(g_1,g_2,g_3)]^2 \,+ \\ &+&\frac{\lambda}{4}\int
dg_1..dg_6 [P_{h_1}\phi(g_1,g_2,g_3)][P_{h_2}\phi(g_3,g_5,g_4)]
[P_{h_3}\phi(g_4,g_2,g_6)][P_{h_4}\phi(g_6,g_5,g_1)]. \end{eqnarray*}
As we have discussed in the general case, we see that the structure of the action is chosen so to reflect the combinatorics of a 3d
triangulation, with four triangles (fields) glued along their edges
(arguments of the field) pairwise, to form a tetrahedron (vertex term) 
and two tetrahedra being glued along their common triangles (kinetic term $\rightarrow$ propagator).  
The quantum theory is given by the partition function, in turn again defined in terms of perturbative expansion in Feynman graphs:
$$Z\,=\,\int d\phi\,e^{-S[\phi]}\,=\,\sum_{\Gamma}\,\frac{\lambda^N}{sym[\Gamma]}\,Z(\Gamma). $$
Therefore, in order to construct explicitly the quantum amplitudes of
the theory for each of its Feynman graphs, we need to identify their
building blocks, i.e. propagator and vertex amplitude. These are to be read
out from the action:
\begin{eqnarray*} S[\phi]=\frac{1}{2}\int
dg_i d\tilde{g}_j\,\phi(g_i)\mathcal{K}(g_i,\tilde{g}_j)\phi(\tilde{g}_j)+\frac{\lambda}{4}\int
dg_{ij}\mathcal{V}(g_{ij})\phi(g_{1j})\phi(g_{2j})\phi(g_{3j})\phi(g_{4j}) \end{eqnarray*}
For the propagator, starting from the kinetic term
$P_g\phi(g_1,g_2,g_3) P_{\bar{g}}\phi(g_1,g_2,g_3)$, and considering
the permutation symmetry, one gets immediately: 
\begin{eqnarray*} \mathcal{P} = \mathcal{K}^{-1} =
    \mathcal{K} = \sum_{\pi}\int dg d\bar{g} \,\,\delta(g_1 g
  \bar{g}^{-1} \tilde{g}_{\pi(1)}^{-1})\delta(g_2 g \bar{g}^{-1}
  \tilde{g}_{\pi(2)}^{-1})\delta(g_3 g \bar{g}^{-1}
  \tilde{g}_{\pi(3)}^{-1}),\end{eqnarray*}
while the vertex is given by:
 \begin{eqnarray*} \mathcal{V} = \int dh_i \,\delta(g_1 h_1
  h_3^{-1} \tilde{g}_1^{-1}) \delta(g_2 h_1 h_4^{-1} \tilde{g}_2^{-1})
  \delta(g_3 h_1 h_2^{-1} \tilde{g}_3^{-1}) \delta(g_4 h_2
  h_4^{-1} \tilde{g}_4^{-1}) \delta(g_5 h_2 h_3^{-1} \tilde{g}_5^{-1})
  \delta(g_6 h_3 h_4^{-1} \tilde{g}_6^{-1})
  \;\;\;\;\;\;\;\;\;\;\;\;\end{eqnarray*}
  
 We see that, as for BF theories in any dimensions, the vertex and propagator are given simply by products of delta functions over the group, represented simply by lines in the diagrams below, with boxes
  representing the integration over the group (following from the requirement of gauge invariance.
\begin{figure}[here]
\setlength{\unitlength}{1cm}
\vspace{-1cm}
\begin{minipage}[t]{6cm}
\begin{picture}(5.0,4.5)
\includegraphics[width=3cm, height=2cm]{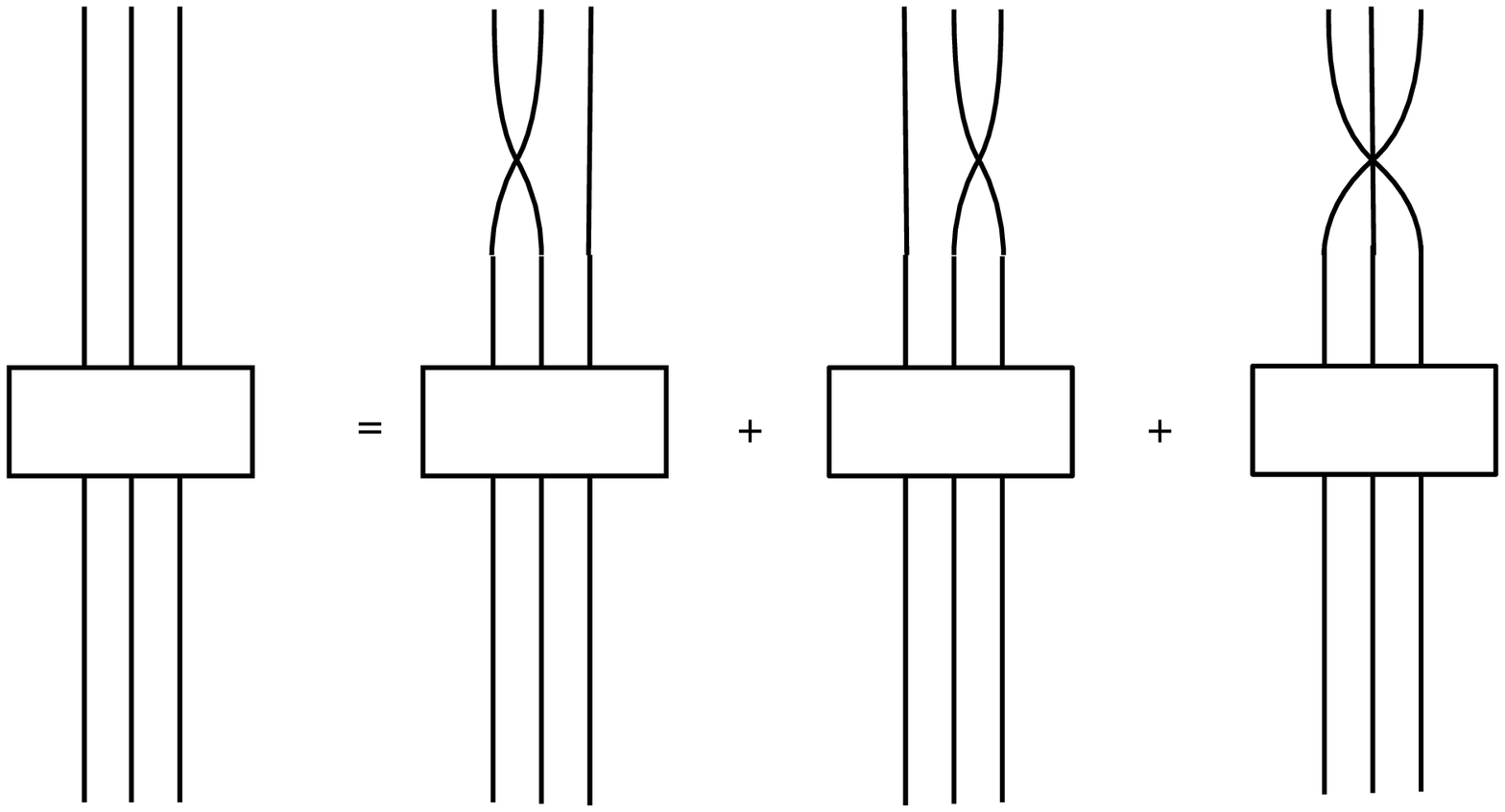}
\end{picture} \par
propagator
\end{minipage}
\hfill
\begin{minipage}[t]{6cm}
\begin{picture}(3.0,4.5)
\includegraphics[width=3cm, height=2.5cm]{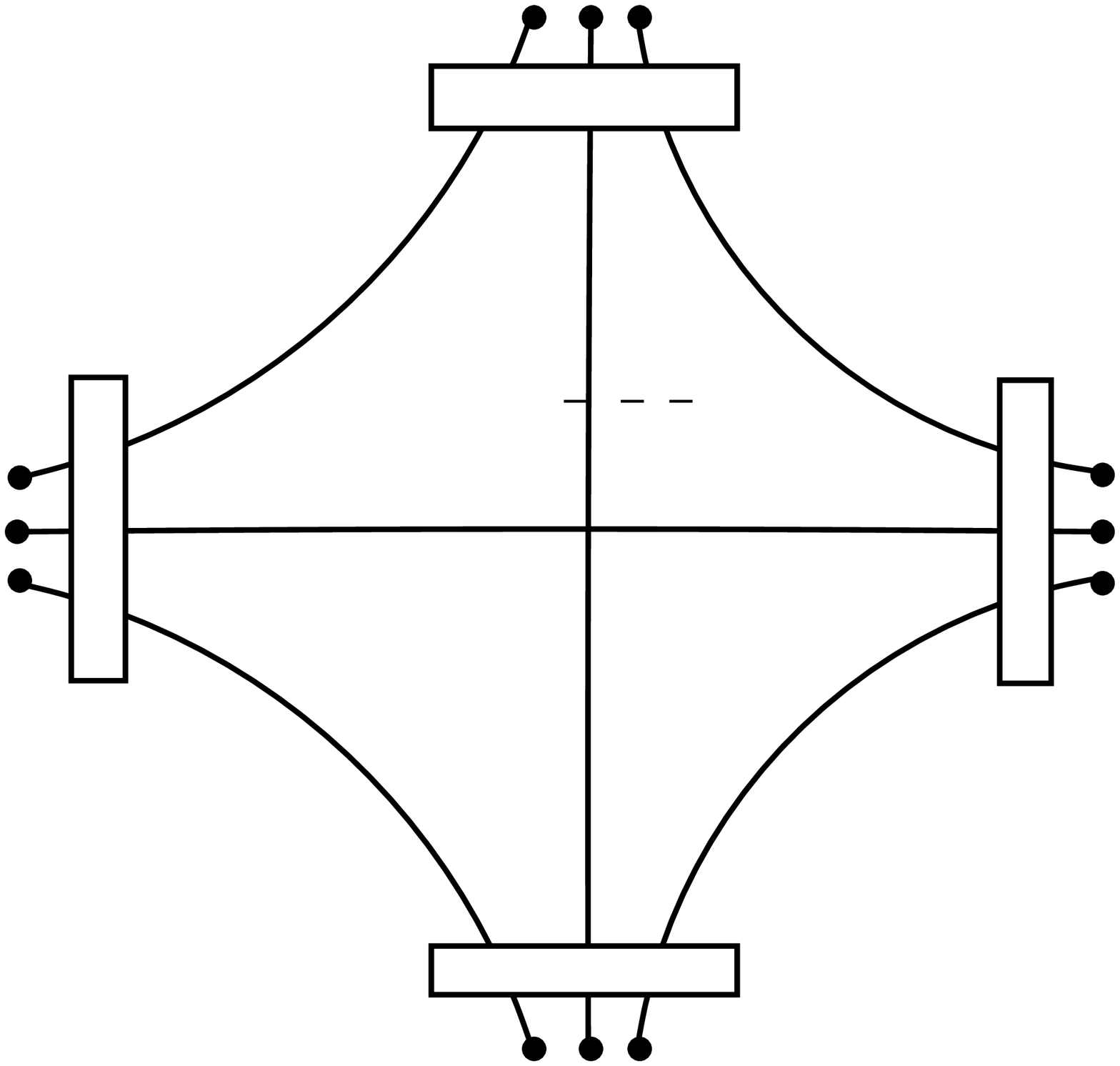}
\end{picture} \par
interaction vertex
\end{minipage}
\end{figure}
The Feynman graphs are obtained as usual by gluing vertices with
propagators. Let us see how these look like. As we explained for the general case, each line in a propagator
goes through several vertices and for closed graphs it comes back to the original point, thus identifying a 2-cell, and these 2-cells, together
with the set of lines in each propagator, and the set
of vertics of the graph, identify a 2-complex. Each of these 2-complexes is dual to a 3d triangulation, with
each vertex corresponding to a tetrahedron, each link to a triangle
and each 2-cell to an edge of the triangulation. 

\begin{figure}[here]
\setlength{\unitlength}{1cm}
\begin{minipage}[t]{3cm}
\begin{picture}(3.0,3.5)
\includegraphics[width=2cm, height=2cm]{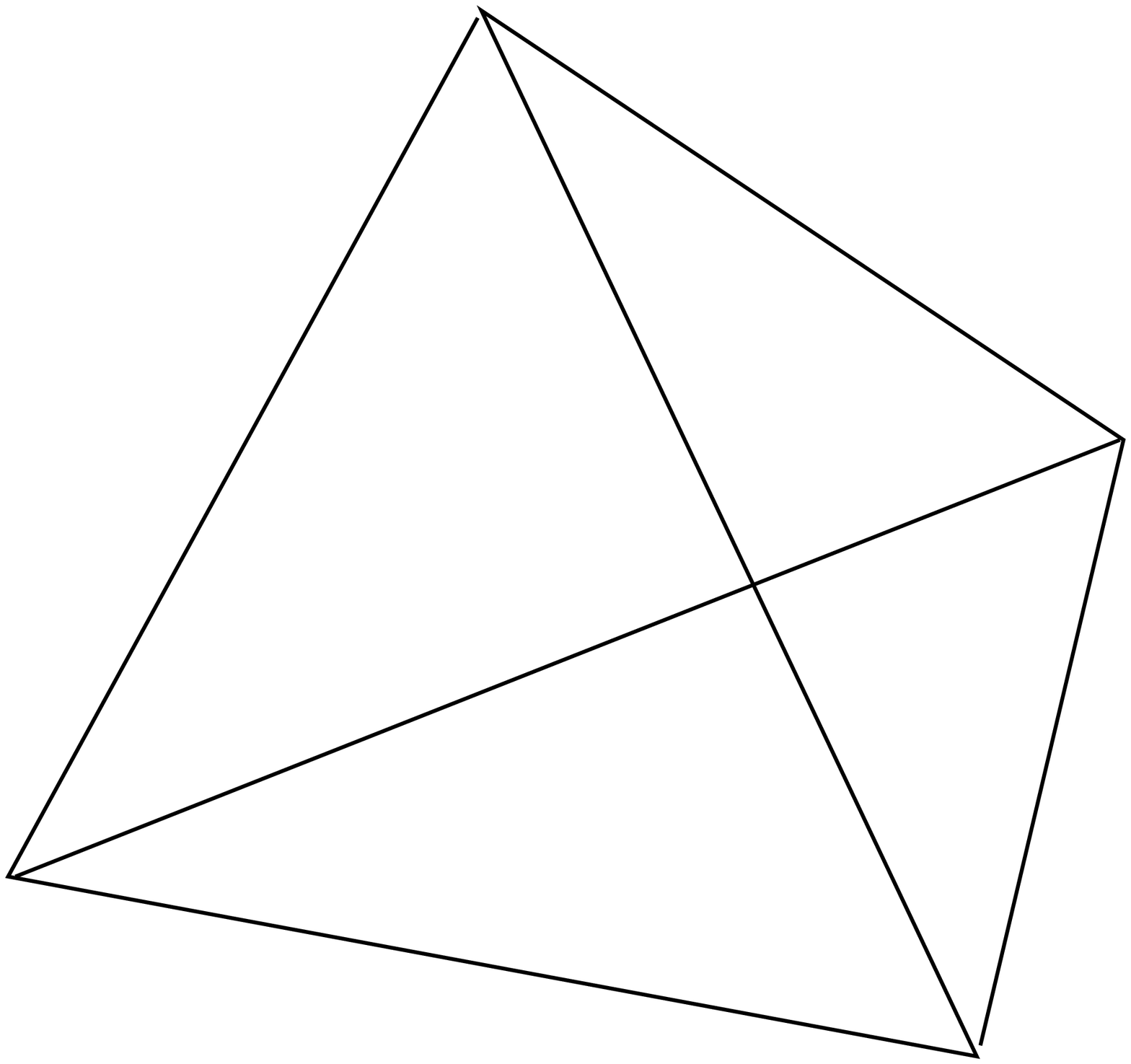}
\end{picture} \par
tetrahedron
\end{minipage}
\hfill
\begin{minipage}[t]{3cm}
\begin{picture}(3.0,3.5)
\includegraphics[width=2cm, height=2cm]{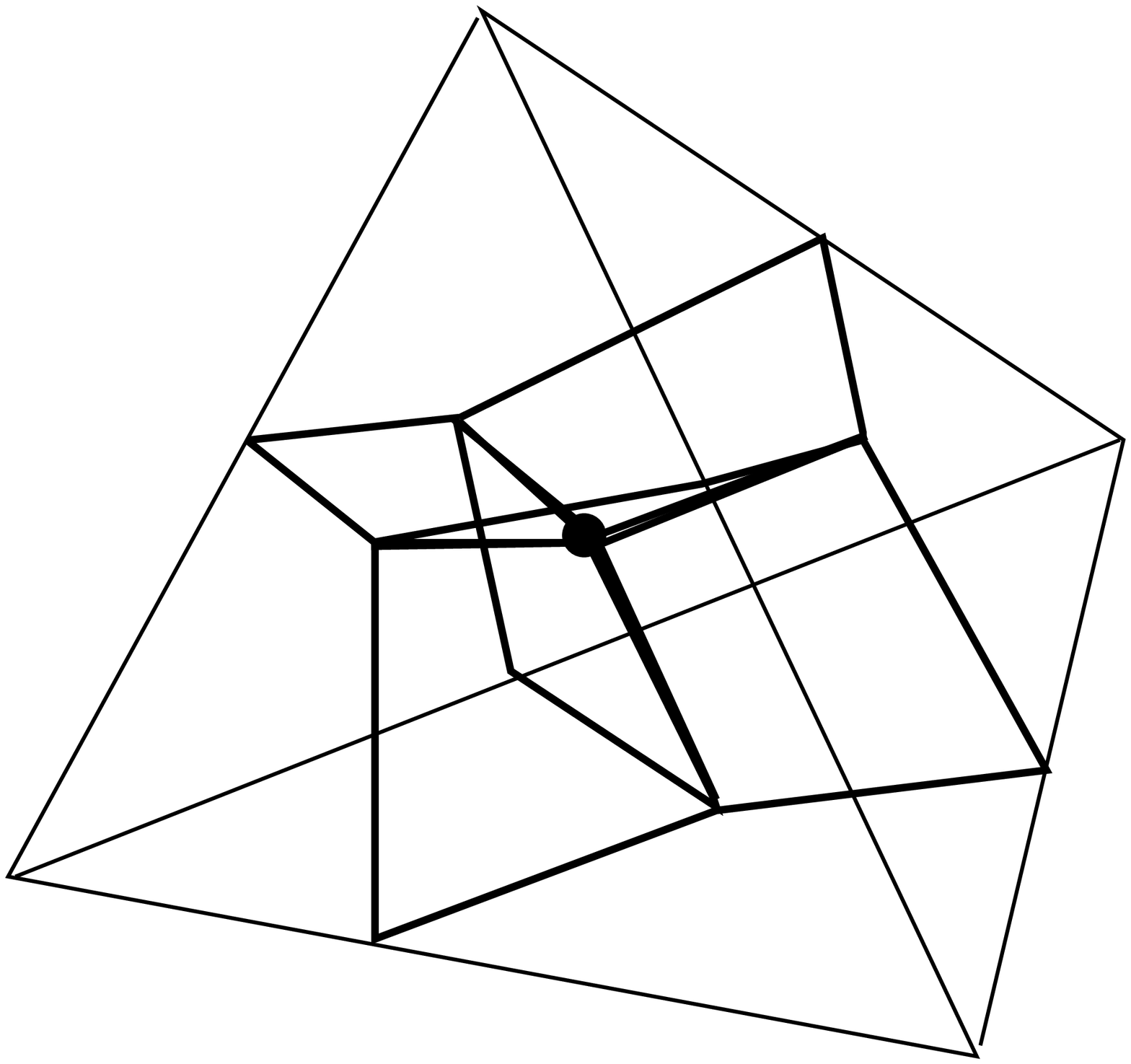}
\end{picture} \par
tetra + dual
\end{minipage}
\hfill
\begin{minipage}[t]{3cm}
\begin{picture}(3.0,3.5)
\includegraphics[width=2cm, height=2cm]{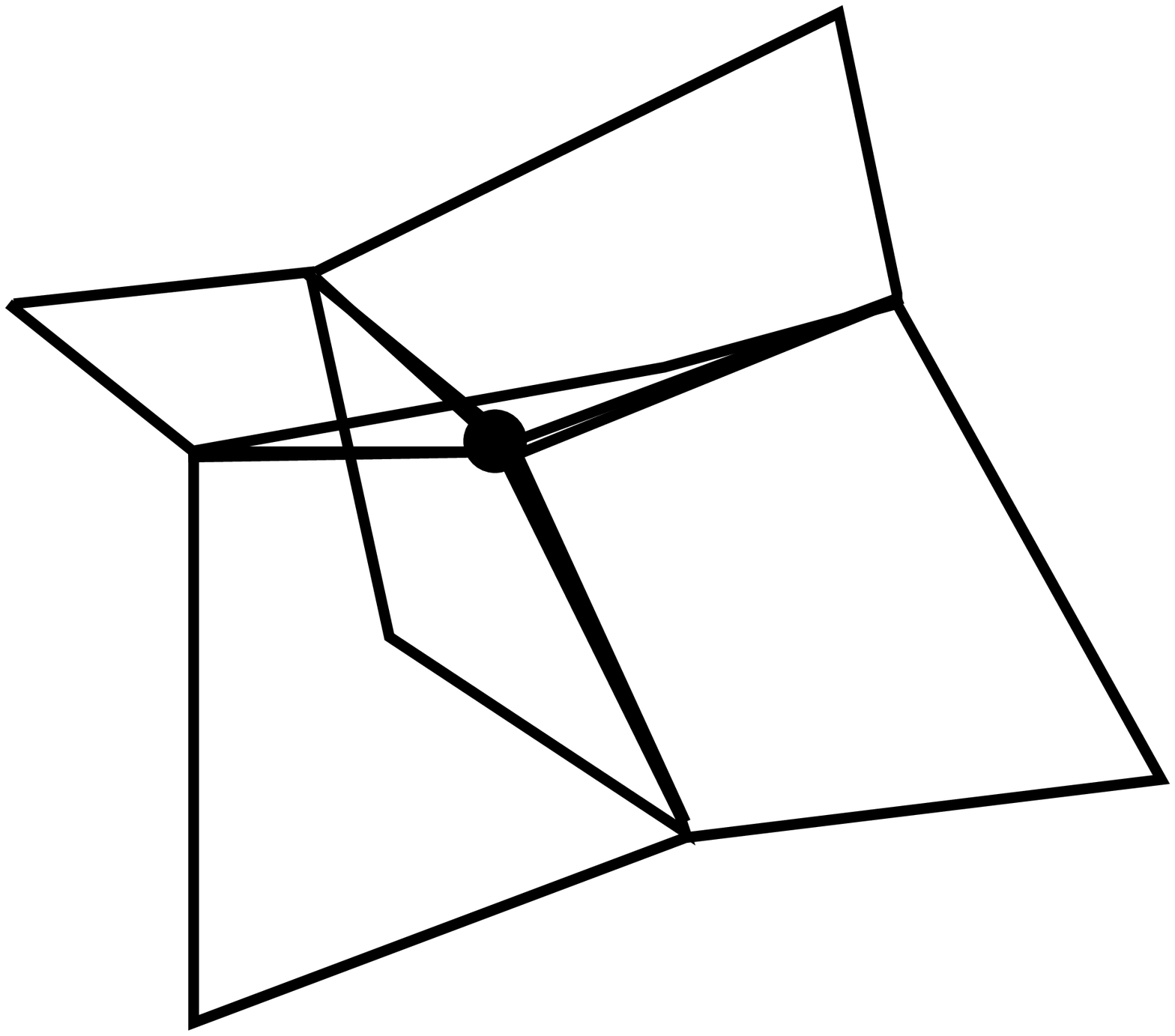}
\end{picture} \par
dual
\end{minipage}
\end{figure}

The sum over Feynman graphs is thus equivalent to a {\bf sum
over 3d triangulations of any topology}, as anticipated in the general case.
Let us now identify the quantum amplitudes for these Feynman graphs. These are obtained the usual way using the above
propagators and vertex amplitudes. In configuration space, where the
variables being integrated over are group elements, the amplitude for
each 2-complex is:
$$
Z(\Gamma)\,=\, \left(\prod_{e\in \Gamma} \int d g_{e}\right)
\,\prod_{f}\,\delta (\prod_{e\in\partial f} g_{e} )
$$
which has the form of a lattice gauge theory partition function with
simple delta function weights for each plaquette (face of the
2-complex) and one connection variable for each edge; the delta
functions constraint the the curvature on any face to be zero, as we
expect from 3d quantum gravity \cite{laurentPRI}.
To have the corresponding expression in momentum space, one expands
the field in modes $\phi(g_1,g_2,g_3)\,=\,\sum_{j_1,j_2,j_3}
\phi^{j_1j_2j_3}_{m_1n_1  m_2n_2
  m_3n_3}\,D^{j_1}_{m_1n_1}(g_1)D^{j_2}_{m_2n_2}(g_2)D^{j_3}_{m_3n_3}(g_3)$, where the $j$'s are irreps of the group $SU(2)$ (the Lorentz group, local gauge group of gravity, for 3d and Riemannian signature) obtaining, for the propagator, vertex and amplitude:

\begin{eqnarray*}
\mathcal{P} &=& \delta_{j_1\tilde{j}_1}\delta_{m_1\tilde{m}_1}
\delta_{j_2\tilde{j}_2} \delta_{m_2\tilde{m}_2}
\delta_{j_3\tilde{j}_3}\delta_{m_3\tilde{m}_3} \\ 
\mathcal{V} &=& \delta_{j_1\tilde{j}_1}\delta_{m_1\tilde{m}_1}
\delta_{j_2\tilde{j}_2} \delta_{m_2\tilde{m}_2}
\delta_{j_3\tilde{j}_3}\delta_{m_3\tilde{m}_3}\delta_{j_4\tilde{j}_4}\delta_{m_4\tilde{m}_4}\delta_{j_5\tilde{j}_5} \delta_{m_5\tilde{m}_5}
\delta_{j_6\tilde{j}_6}\delta_{m_6\tilde{m}_6} \left\{
\begin{array}{ccc} 
j_1 &j_2 &j_3
\\ j_4 &j_5 &j_6 
\end{array}\right\}
\\ Z(\Gamma)&=&\left(\prod_{f}\,\sum_{j_{f}}\right)\,\prod_{f}\Delta_{j_{f}}\prod_{v}\, \left\{ \begin{array}{ccc} 
j_1 &j_2 &j_3
\\ j_4 &j_5 &j_6 
\end{array}\right\} \end{eqnarray*}  
where $\Delta_j$ is the dimension of the representation $j$ and for
each vertex of the 2-complex we have a so-called $6j-symbol$ , i.e. a
scalar function of the 6 representations meeting at that vertex. The
amplitude for each 2-complex is given then by a spin foam model, the
Ponzano-Regge model for 3d gravity without cosmological constant,
about which a lot more is known \cite{laurentPRI}. This amplitude, after gauge fixing, gives a well-defined topological invariant of 3-manifolds, as one expects from 3d quantum gravity, and as such it is invariant under choice of triangulation. This means that it evaluates to the same number for any triangulation (2-complex) for given topology, so that the only dynamical degrees of freedom in the theory are indeed the topological.
The full theory is then defined as we said by the sum over all Feynman
graphs weighted by the above amplitudes:
$$ Z =\sum_{\Gamma}\,\frac{\lambda^N}{sym[\Gamma]}\,\left(\prod_{f}\,\sum_{j_{f}}\right)\,\prod_{f}\Delta_{j_{f}}\prod_{v}\,\left\{ \begin{array}{ccc} 
j_1 &j_2 &j_3
\\ j_4 &j_5 &j_6 
\end{array}\right\}_v. $$
This gives a rigorous (after gauge-fixing of the symmetries of the theory \cite{laurentdiffeo} the above quantity is well-defined, even in the Lorentzian case, in spite of the non-compactness of the Lorentz group used and of the infinite dimensionality of the irreps used) and un-ambiguous (in the sense that every single element in the above formula has a known closed expression and therefore it can be computed exactly at least in principle) realisation, in purely algebraic
and combinatorial terms, of the sum over both geometries and
topologies, i.e. of the third quantization idea, in the 3-dimensional case. Issues about interpretation and about the convergence of the sum over complexes (but see \cite{sumtop}) of course remain, but we see a definite progress at least fro what concerns the definition of the amplitude and measure for given 2-complex (i.e. given spacetime) with respect to the continuum path integral.  

Strikingly, group field theory models for quantum gravity in 4
spacetime dimensions, that seem to have many of the right properties
we seek, can be obtained by a very simple modification of the
3-dimensional model \cite{review,alex}. Motivated by the classical
formulation of gravity as a constrained BF theory \cite{review, alex},
one first generalises the above field to a 4-valent one with arguments
living in the 4-dimensional Lorentz group $SO(3,1)$, modifying the
combinatorial structure of the above action to mimic the combinatorics
of a 4-simplex in the interaction term, with 5 tetrahedra (fields)
glued along triangles, and then simply imposes a restriction on the
arguments of the field to live in the homogeneous space
$SO(3,1)/SO(3)\simeq H^3$, i.e. on the upper sheet of the timelike
hyperboloid in Minkowski space. The resulting Feynman graphs expansion
produces quantum amplitudes for them given by the Barrett-Crane model
\cite{BC}, that has been recently the focus of much work for which we
refer to the literature \cite{review, alex}.   

\section{Assorted questions for the present, but especially for the future}
We have outlined the general formalism of group field theories and the
picture of a third quantized simplicial spacetime they suggest. We
have organised this outline and presented these models from the perspective that sees GFTs as a candidate to and a proposal for a fundamental formulation of a theory of
quantum gravity, as opposed to just a tool to be used to produce for
example spin foam models, that is the way they have been used
up to now. We have tried to highlight the features of the formalism
and of the resulting picture that we find most appealing and
fascinating. However, it is probably transparent that there is and
there should be much more in the theory than what we have
described. As a matter of fact, there is certainly in the GFTs much
more than it is presently known, which is basically only their action,
their perturbative expansion, and a few properties of their Feynman
amplitudes, i.e. spin foam models. To tell the whole truth, even many of
the details of the general picture we have presented are only
tentative and there remains lots to be understood about
them. Therefore we want to conclude by posing a (limited) set of
assorted questions regarding GFTs and that future work should answer,
if GFTs are to be taken seriously as candidates for a fundamental
formulation of third quantized simplicial gravity, and thus of
non-perturbative quantum gravity.

\begin{itemize}
\item {\bf What about the classical theory?}
The description of the classical simplicial superspace that we have
briefly described is to be investigated further as the validity of the
variables that GFTs use to describe classical geometry is not solidly
established yet. Also, even given this for granted, not much is known
about the classical theory behind the GFT action. Some work is in
progress \cite{instantons}, but it is fair to say that we do not have
a good understanding of the physical meaning of the classical equations of
motion of the theory nor we know enough solutions of them. As for the physical menaing, we stress again that the {\it classical} GFT dynamics should already encode the canonical {\it quantum gravity} dynamics in full, as it happens in ordinary quantum field theories (where the classical field equations are the quantum Schroedinger equation for the one-particle theory). Also, in analogy
to the continuum third quantization setting, one would expect them to
describe a modification of the Wheeler-DeWitt equation, in a simplicial setting, due to the
presence of topology change; however, unlike the continuum case it is
not easy, and maybe not even possible, to distinguish (the simplicial
analogue of) a Wheeler-DeWitt operator and a topology changing
term as distinct contributions to the equation, due to the local
nature of the description of superspace that we have in GFTs. As for obtaining solutions of the classical equations of motion,
the non-local nature of the equations of motion (coming from that of
the action) makes solving them highly non-trivial even in lower
dimensions. The form of the action in the generalised GFT formalism
\cite{generalised} may probably facilitate this task somehow, due to a
greater similarity with usual scalar quantum field theories. Similarly
for the Hamiltonian analysis of the theory, not yet performed  and
that is quite non-trivial in the usual formalism, but may be easier to
carry out in a covariant fashion thanks to the proper time variables
in the generalised formalism. In this regard, the fact extablished in \cite{generalised} that usual GFTs are the Static Ultra-local (SUL), and un-oriented, limit of well-defined generalised GFTs may help in that one may first tackle with standard methods the Hamiltonian analysis of such generalised models and then study the SUL limit of the corresponding Hamiltonian structures.

\item{\bf Fock space and +ve/-ve energy states?}
The picture of creation and annihilation of quanta of simplicial
geometry, the construction of quantum states of the field theory, and
the Fock structure of the corresponding space of states built out of
the 3rd quantized vacuum may be appealing and geometrically
fascinating, but it is at present mostly based on intuition. A
rigorous construction needs to be carried out, starting from a proper
hamiltonian analysis, and a convincing identification of creation and
annihilation operators from the mode expansion of the field. Having at
hand the hamiltonian form and the harmonic decomposition of the field
itself in modes (resulting from the harmonic analysis on group
manifolds), one should investigate the quantum gravity analogue of the
notion of positive and negative
energy (particle/anti-particle) states of QFT. This has probably to do
with the opposite orientations one can assign to the fundamental
building blocks of our quantum simplicial spacetime, so maybe it is again best investigated in the generalised GFT formalism \cite{generalised}, in which the orientation data play a crucial role.  
\item {\bf What is $\lambda$?}
In
  \cite{laurentgft}, one consistent and convincing interpretation of the GFT coupling constant has been given: it was shown that a suitable
  power of the coupling constant can be interpreted as the parameter
  governing the sum over topologies in the perturbative expansion in
  Feynman graphs. This result was based on the analysis of the Schwinger-Dyson equations of the n-dimensional GFT. More work is now needed to expand on these result and elucidate all its implications. Another interpretation of the GFT coupling constant, although not as well extablished is suggested by the work \cite{DP-P}. If one considers only one fixed representation of the
  Lorentz group, thus reducing GFTs to
  simple tensor extensions of matrix models, then the coupling constant
  enters in the resulting amplitudes as the exponential of a cosmological
  constant, and the perturbative expansion is
  interpreted as an expansion in the \lq\lq size\rq\rq of spacetime \cite{DP-P};
  this is beautiful, but one must check whether a
  similar picture applies for the full GFT with
  no restriction on the representations.  Assuming this can be extablished, more work is then needed to prove the compatibility of the two
  interpretations. This last point has
  important consequences, in that it may lead to a simplicial and
  exact realisation of the old idea, put forward in the context of
  continuum formulations of 3rd quantised gravity
  \cite{banks}, of a connection
  between cosmological constant and wormholes, i.e. topology change,
  in quantum gravity.

\item{\bf Where are the diffeos? actually, where are all the continuum symmetries?}
Another crucial point that needs to be addressed is a deeper
understanding of GFT symmetries; in fact, already in the 3-dimensional
case it is known \cite{laurentPRI} that the amplitudes generated by the GFT
possess symmetries that are not immediately identified as symmetries
of the GFT action.
The most important
symmetry of a gravity theory is diffeomorphism symmetry. One could argue that GFTs are manifestly diffeomorphism invariant in the sense that
there is no structure in the theory corresponding to a continuum
spacetime, but one should identify a discrete symmetry
corresponding to diffeos in the
appropriate continuum approximation of the theory. Since any
continuum metric theory that is invariant under diffeomorphisms
reproduces Einstein gravity (plus higher derivative terms), the identification of
diffeomorphism symmetry in GFTs is crucial, it that it would support the idea that they possess a continuum
approximation given by General Relativity. It was argued in
\cite{laurentgft} that
diffeomorphisms are the origin of loop divergences in spin foam
models, that in turn are just Feynman amplitudes for the GFT. This needs to be investigated. Another possibility to be investigated is that diffeomophisms
originate from a non-trivial renormalisation group acting on the
parameter space of GFTs. 

\item {\bf A GFT 2-point functions zoo? where is causality?}
In ordinary quantum field theory, one can define different types of
N-point functions or transition amplitudes, with different uses and
meanings; is this the case also for GFTs, and more generally for
Quantum Gravity? if so, what is their respective use and
interpretation? The difference between various N-point functions in
QFT is in their different causal properties, so this question is
related to the more general issue of causality in GFTs. Where is
causality? How to implement causality restrictions? Recent work
\cite{generalised} seems to suggest that these issues can be dealt
with satisfactorily, but much more work is certainly needed.  

\item {\bf What is the exact relation with the canonical theory?}
A canonical theory based on Lorentz spin networks, adapted to a
simplicial spacetime, and thus a kind of covariant discretization of
loop quantum gravity \cite{carlobook}, has to be given by a subsector
of the GFT formalism. The reduction to this subsector as well as the
properties of the resulting theory are not yet fully understood. A
canonical theory of quantum gravity needs a spacetime topology of product type, i.e. $\Sigma \times \mathbb{R}$, so that the only non-trivial topology can be in the topology of space $\Sigma$.
and a positive definite inner product between quantum states. A
precise and well-posed definition of such an inner product and a way to
reduce to trivial topology in the perturbative expansion of the GFT
was proposed in \cite{laurentgft}, as we discussed: it is given by the perturbative
evaluation of the expectation value of appropriate spin network
observables in a {\it tree level} truncation, that indeed generates
only 2-complexes with trivial topology. The consequences of this
proposal, that would amount to a complete definition of the canonical
theory corresponding to the GTFs, need to be investigated in
detail. It may be possible for example to extract a definition of an
Hamiltonian constraint operator from the so-defined inner product, to
study its properties, and to compare it with those proposed in the
loop quantum gravity approach. Even when this has been done, it would remain to investigate the relation between covariant spin network structures based on the full Lorentz group, and the loop quantum gravity ones based on $SU(2)$, and to check how much of the many mathematical results obtained on the kinematical Hilbert space of $SU(2)$ spin networks can be reproduced for the covariant ones. 

\item {\bf How to include matter?}
The inclusion and the correct description of matter fields at the
group field theory level is of course of crucial importance. Work on
this has started only recently \cite{kirill, us, gftmatter} for the
3-dimensional case, with very interesting results. The idea pursued
there was that one could perform a 3rd quantization of gravity and a
2nd quantization of matter fields in one stroke, thus writing down a
coupled GFT action for both gravity and matter fields that would
produce, in perturbative expansion, a sum over simplicial complexes
with dynamical geometry (quantum gravity histories) together with
Feynman graphs for the matter fields living on the simplicial
complexes (histories for the matter fields). The whole description of
the coupled system would thus be purely algebraic and
combinatorial. Indeed, this can be realised consistently
for any type of matter field \cite{gftmatter}. However, in 3d life is
made easier by the topological nature of gravity and by the fact that
one can describe matter as a topological defect. The difficult task
that lies ahead is to extend these results to 4 dimensions. In this
much more difficult context, some work i currently in progress
regarding the coupling of gauge fields to quantum gravity at the GFT
level \cite{usgauge}.

\item {\bf Does the GFT perturbation theory make sense?}
Even if the only thing we know about GFTs is basically their
perturbative expansion in Feynman graphs, strangely enough we do not
know for sure if this perturbative expansion makes sense. Most likely
the perturbative series is not convergent, but this is not too bad, as
it is what happens in ordinary field theories. One would expect
(better, hope) it to be an asymptotic series to a non-perturbatively
defined function, but this has been realised up to now only in a
specific model in 3d \cite{sumtop}, and more work is needed for what
concerns other models especially in 4d. Let us recall that the
perturbative GFT expansion entails a sum over topologies, so that
gaining control over it is a mathematically highly non-trivial issue
with very important physical consequences. 
  
\item {\bf How to relate to the other approaches to quantum gravity?}
Even if one is optimistic and buys the picture of GFTs as a general
framework for quantum gravity encompassing other approaches, or at least the main ingredients of other approaches, the links
with these approaches need to be investigated in detail to start
really believing the picture. For example, to obtain a clear link with
simplicial approaches to quantum gravity, one needs first to construct
a GFT that has Feynman amplitudes given by the exponential of the
Regge action for the corresponding
simplicial complex, probably building on the results of
\cite{generalised}. Then one would be left to investigate the
properties of the measure in front of the exponential, to be compared
with those used in Regge calculus, and to find a nice procedure for
reducing the model to involve only equilateral triangulations and to
admit a slicing structure, so to compare it with dynamical
triangulations models. And this would be just a start.

\item {\bf What about doing some physics?} The ultimate aim is of
  course to have a consistent framework for describing quantum gravity
  effects and obtain predictions that can be compared to
  experiments. This may be seen as far-fetched at present, and maybe
  it is, but a consistent coupling of matter fields at the GFT level,
  a better understanding of its semiclassical states and of
  perturbations around them, and a better control over the continuum
  approximation of the GFT structures, all achievable targets for
  current and near future studies, may bring even this ultimate aim
  within our reach not too far from now.

\end{itemize}

\section*{Acknowledgments}
It is my pleasure to thank K. Krasnov, E. Livine, K. Noui, A. Perez,
C. Rovelli, J. Ryan and especially L Freidel for many discussions, clarifications and
suggestions; I thank an anonymous referee for his/her helpful comments
and criticisms; I also thank the organizers of the Blaubeuren Workshop on
\lq\lq Mathematical and Physical Aspects of Quantum Gravity\rq\rq for
their invitation to participate.


\begin{thebibliography}{99}
\bibitem{introGFTshort} D. Oriti, in the Proceedings of the 4th Meeting on Constrained Dynamics and Quantum Gravity, Cala Gonone, Italy (2005); gr-qc/0512048;
\bibitem{laurentgft} L. Freidel, hep-th/0505016;
\bibitem{GFTbook} D. Oriti, in {\it Approaches to Quantum Gravity -
  toward a new understanding of space, time and matter}, D. Oriti ed., Cambridge University Press (2006);
\bibitem{hartle} J. Hartle, in Les Houches Sum.Sch.1992:0285-480, gr-qc/9304006;
\bibitem{canodiffeo} C. Isham, K. Kuchar, Annals Phys. 164, 316 (1985);
\bibitem{euclQG} G. Gibbons and S. Hawking, eds., {\it Euclidean Quantum Gravity}, World Scientific, Singapore (1993);
\bibitem{fay} F. Dowker, in {\it The future of theoretical physics and cosmology}, 436-452, Cambridge University Press (2002), gr-qc/0206020;
\bibitem{horo} G. Horowitz, Class. Quant. Grav. 8, 587-602 (1991);
\bibitem{fayrafael} F. Dowker, R. Sorkin, Class. Quant. Grav. 15, 1153-1167 (1998), gr-qc/9609064;
\bibitem{greene} P. Anspinwall, B. Grene, D. Morrison, Nucl. Phys. B 416, 414-480 (1994), hep-th/9309097;
\bibitem{banks} T. Banks, Nucl. Phys. B 309, 493 (1988);
\bibitem{coleman} S. Coleman, Nucl. Phys. B 310, 643 (1988);
\bibitem{isham} C. Isham, gr-qc/9510063;
\bibitem{sorkin} R. Sorkin, Int. J. Theor. Phys. 30, 923-948 (1991);
\bibitem{ruth} H. Hamber, R. M. Williams,  Nucl. Phys. B 415, 463-496 (1994), hep-th/9308099;
\bibitem{carlobook} C. Rovelli, {\it Quantum Gravity}, Cambridge
  University Press (2004); 
\bibitem{review} D. Oriti, Rept. Prog. Phys. 64, 1489 (2001), gr-qc/0106091;
\bibitem{alex} A. Perez, Class. Quant. Grav. 20, R43 (2003), gr-qc/0301113;
\bibitem{CS} F. Dowker, gr-qc/0508109;
\bibitem{thesis} D. Oriti, {\it Spin foam models of a quantum
  spacetime}, PhD thesis, University of Cambridge (2003), gr-qc/0311066;
\bibitem{giddingsstrominger} S. B. Giddings, A. Strominger,
  Nucl. Phys. B 321, 481 (1989);
\bibitem{guigan} M. McGuigan, Phys. Rev. D 38, 3031-3051 (1988);
\bibitem{matrix} A. Morozov, hep-th/0502010;
\bibitem{DT} J. Ambjorn, J. Jurkiewicz, R. Loll, Phys. Rev. D 72,
  064014 (2005), hep-th/0505154;
\bibitem{carlomike} M. Reisenberger, C. Rovelli,  Class. Quant. Grav. 18, 121-140 (2001), gr-qc/0002095; 
\bibitem{generalised} D. Oriti, gr-qc/0512069;
\bibitem{bb} J. Baez, J. Barrett, Adv. Theor. Math. Phys. 3, 815-850 (1999), gr-qc/9903060;
\bibitem{DP-P} R. De Pietri, C. Petronio, J. Math. Phys. 41, 6671-6688 (2000), gr-qc/0004045;
\bibitem{mikovic} A. Mikovic, Class. Quant. Grav. 18, 2827-2850 (2001), gr-qc/0102110;
\bibitem{instantons} A. Baratin, L. Freidel, E. Livine, in preparation;
 \bibitem{laurentPRI} L. Freidel, D. Louapre, Class. Quant. Grav. 21,
  5685 (2004), hep-th/0401076;
\bibitem{boulatov} D. Boulatov, Mod.Phys.Lett. A7 (1992) 1629-1646,
  hep-th/9202074; 
\bibitem{BC} J. W. Barrett, L. Crane, Class.Quant.Grav. 17 (2000)
  3101-3118, gr-qc/9904025;
  \bibitem{kirill} K. Krasnov, hep-th/0505174;
\bibitem{us} L. Freidel, D. Oriti, J. Ryan, gr-qc/0506067; 
\bibitem{gftmatter} D. Oriti, J. Ryan, gr-qc/0602010;
\bibitem{usgauge} R. Oeckl, D. Oriti, J. Ryan, in preparation;
\bibitem{sumtop} L. Freidel, D. Louapre, Phys. Rev. D 68, 104004 (2003), hep-th/0211026.
\end{thebibliography}
\end{document}